\documentclass[12pt]{article}
\usepackage[a4paper,top=2.5 cm,bottom=2.5cm,left=2.5cm,right=2.5cm]{geometry}

\usepackage[utf8]{inputenc}				
\usepackage{mdframed}					
\usepackage{graphicx}					

\usepackage{caption}
\usepackage{subcaption}

\graphicspath{{figures/}}
\usepackage{xspace}
\usepackage{amsmath}

\usepackage{amsthm}
\theoremstyle{remark}
\newtheorem*{remark}{Remark}

\usepackage{mathtools}
\usepackage{enumitem}
\usepackage[super]{nth}

\usepackage[backend=bibtex, sorting=none, style=numeric, citestyle=numeric]{biblatex}

\DeclareNameAlias{sortname}{last-first}

\usepackage{hyperref}

\mathtoolsset{showonlyrefs}

\usepackage{comment}

\usepackage{float}

\usepackage{enumitem}

\usepackage{color}

\usepackage{physics}

\usepackage{amssymb}
\usepackage{upgreek}

\newcommand{\defeq}{\vcentcolon=}

\newcommand{\bbR}{\ensuremath\mathbb{R}} 

\newcommand{\vect}[1]{%
	{\boldsymbol{\mathbf{%
				\mathit{#1}
	}}}
}

\newcommand{\ctens}[1]{%
	{\boldsymbol{\mathbf{%
				\mathit{#1}
	}}}
}

\newcommand{\SDvect}[1]{%
{\mathbf{\mathbf{%
            \mathbf{#1}
    }}}
}

\newcommand{\SDtens}[1]{%
{{
{\mathbf{%
            \mathbf{#1}
    }}}}
}

\newcommand{\ctensd}[1]{\mathsf{{#1}}}

\usepackage{authblk}
	
\usepackage{color}
	
	\title{A time multiscale decomposition in cyclic elasto-plasticity}
	
	\author[,1,3]{Angelo Pasquale\thanks{Corresponding author}}
	\author[1]{Sebastian Rodriguez}
	\author[2]{Khanh Nguyen}
	\author[3,4]{Amine Ammar}
	\author[1,4]{Francisco Chinesta}
	
	\affil[1]{ESI Group Chair @ PIMM Lab, ENSAM Institute of Technology, 151 Boulevard de l'Hôpital, F-75013, Paris, France, \href{mailto:angelo.pasquale@ensam.eu}{angelo.pasquale@ensam.eu},
		\href{mailto:sebastian.rodriguez_iturra@ensam.eu}{sebastian.rodriguez\_iturra@ensam.eu},
		\href{mailto:francisco.chinesta@ensam.eu}{francisco.chinesta@ensam.eu}\vspace{0.5cm}}
	\affil[2]{Escuela Técnica Superior de Ingeniería Aeronáutica y del Espacio, Universidad Politécnica de Madrid, Pza. Cardenal Cisneros, 28040, Madrid, Spain, \href{mailto:khanhnguyen.gia@upm.es}{khanhnguyen.gia@upm.es}\vspace{0.5cm}}
	\affil[3]{ESI Group Chair @ LAMPA Lab, ENSAM Institute of Technology, 2 Boulevard du Ronceray BP 93525, 49035 Angers cedex 01, France, \href{mailto:angelo.pasquale@ensam.eu}{angelo.pasquale@ensam.eu},
		\href{mailto:amine.ammar@ensam.eu}{amine.ammar@ensam.eu}\vspace{0.5cm}}
	\affil[4]{CNRS@CREATE LTD, 1 Create Way, \#08-01 CREATE Tower, Singapore 138602, Singapore}
	
	\date{}
	
	\makeatletter
	\newcommand*{\toccontents}{\@starttoc{toc}}
	\makeatother

\usepackage{bbm}
\usepackage{tikz,pgfplots}

\usetikzlibrary{decorations.fractals}

\pgfplotsset{compat=newest}
\usetikzlibrary{plotmarks}
\usetikzlibrary{arrows}
\usepgfplotslibrary{patchplots}
\usepackage{grffile}
\pgfplotsset{plot coordinates/math parser=true}
\usepackage{standalone}

\usetikzlibrary{arrows.meta,graphs,decorations.pathmorphing,decorations.markings} 

\usepackage{titlesec}
\usepackage[titletoc,toc,title]{appendix}

\DeclareMathSymbol{\mlq}{\mathord}{operators}{'134}
\DeclareMathSymbol{\mrq}{\mathord}{operators}{'42}

\usepackage{parskip}

\usepackage{algpseudocode,algorithm,algorithmicx}

\algrenewcommand\algorithmicrequire{\textbf{Require:}}
\algrenewcommand\algorithmicensure{\textbf{Postcondition:}}
\algnewcommand\algto{\textbf{ to }}

\usepackage{bbold} 

\begin{document}						

		
		\maketitle
		
	\begin{abstract}
            For the numerical simulation of time-dependent problems, recent works suggest the use of a time marching scheme based on a tensorial decomposition of the time axis. This time-separated representation is straightforwardly introduced in the framework of the Proper Generalized Decomposition (PGD). The time coordinate is transformed into a multi-dimensional time through new separated coordinates, the micro and the macro times. 
            From a physical viewpoint, the time evolution of all the quantities involved in the problem can be followed along two time scales, the fast one (micro-scale) and the slow one (macro-scale). In this paper, the method is applied to compute the quasi-static response of an elasto-plastic structure under cyclic loadings. The study shows the existence of a physically consistent temporal decomposition in computational cyclic plasticity. Such micro-macro characterization may be particularly appealing in high-cycle loading analyses, such as aging and fatigue, addressed in a future work in progress.\\

	\end{abstract}
	
	
	\section{Introduction}
    Despite the availability of high performance computing platforms, the numerical solution of complex, time-dependent nonlinear problems encountered in solid mechanics still nowadays remains a cumbersome challenge. This happens, for instance, when dealing with problems defined over very large time intervals (e.g., fatigue, aging, dynamics with loadings involving multiple characteristic times) and where the system response must encompass the different time scales present in the model \cite{cyclic-plasticity, cyclic-plasticity-1, cyclic-plasticity-2, cyclic-plasticity-3}. This scenario makes difficult the use of a numerical simulation based on classical time marching schemes. 
    
    In recent works, when addressing a problem with long-term horizon and involving different times scales, it has been proposed to separate the time by using a multi-time separated representation \cite{pgd-multiscale,pgd-multiscale-pu,pgd-multiscale-1}. Neglecting the time derivatives, a quasi-static (nonlinear) problem can be expressed as
    \begin{equation}
        \label{eq:generic-problem}
        \mathcal{L} (u (\vect{x}, t)) = f(\vect{x}, t)
    \end{equation}
    where $\mathcal{L} (\bullet)$ refers to a generic nonlinear differential operator involving the derivatives in space. Here the time dependence is associated to the loading $f(\vect{x}, t)$.

    If the time coordinate must resolve very fine details, while it must cover a long time interval, as usually encountered in fatigue analyses, then, without assuming scales separation, one could define a fast and a slow time (other intermediate scales could be also considered), $\tau$ and $T$ respectively. As shown in \cite{pgd-multiscale,pgd-multiscale-1,pgd-multiscale-2}, using the Proper Generalized Decomposition (PGD), the solution of \eqref{eq:generic-problem} can efficiently be computed in the separated space/multi-time form as soon as $\mathcal{L}$ is a linear operator.
    
    In case of nonlinearities, the nonlinear operator $\mathcal{L}$ can be decomposed in a linear part $\mathcal{L}_{\text{l}}$ and a nonlinear one $\mathcal{L}_{\text{nl}}$. In such a way, problem \eqref{eq:generic-problem} can be rewritten as \eqref{eq:generic-problem-l-nl}
    \begin{equation}
        \label{eq:generic-problem-l-nl}
        \mathcal{L}_{\text{l}} (u (\vect{x}, t)) = f(\vect{x}, t) - \mathcal{L}_{\text{nl}} (u (\vect{x}, t)),
    \end{equation}
    which is easily linearized, for instance as
    \begin{equation}
        \label{eq:generic-problem-linearized}
        \mathcal{L}_{\text{l}} (u^{(l)} (\vect{x}, t)) = f(\vect{x}, t) - \mathcal{L}_{\text{nl}} (u^{(l-1)} (\vect{x}, t)),
    \end{equation}
    where the superscript $(l)$ refers to the nonlinear iteration. The solution of the linearized problem can be computed using a full space/micro-time/macro-time decomposition \cite{pgd-multiscale,pgd-multiscale-1}
    \begin{equation}
        u^{(l)}(\vect{x}, t) \approx \sum_{k} U^{\vect{x}}_k(\vect{x}) U^\tau_k(\tau) U^T_k(T),
    \end{equation}
    or starting with a space/time separation and imposing a further multi-time decomposition for the computation of the time function (micro/macro time sub-modes) \cite{pgd-multiscale-2} 
    \begin{equation}
        \label{eq:space-multimodes-0}
        u^{(l)}(\vect{x}, t) \approx \sum_{k} U^{\vect{x}}_k(\vect{x}) U^t_k(t) \approx \sum_{k} U^{\vect{x}}_k(\vect{x}) \sum_j U^\tau_{k, j}(\tau) U^T_{k, j}(T).
    \end{equation}

    The aim of this paper is computing such a time multiscale representation when dealing with cyclic elasto-plasticity, where the non-linear term depends on the past history, that is
    \begin{equation}
    	\mathcal{L}_{\text{nl}}(u^{(l-1)}(\vect{x}, t)) = \int_0^t \mathcal{L}_{\text{nl}}(u^{(l-1)}(\vect{x}, s)) \dd{s}.
    \end{equation} 

	The present paper is not loooking for computational effectiveness, but only to prove that the time decomposition operates and correctly performs when assuming a time-story nonlinear term, as plasticity promotes. 
	
	A paper in progress will address the computational effectiveness by combining discretization and machine learning within the time-separated representation here proposed. 

	The paper is structured as follows. Section \ref{sec:model-problem} presents the problem statement in its strong and weak forms. Section \ref{sec:PGD-based-time-multiscale} recasts the problem in the numerical framework of the Proper Generalized Decomposition, starting from the space and time separation and then addressing the multi-time separation. Section \ref{sec:application-case} shows the results on load-unload tensile tests. Section \ref{sec:conclusion} gives conclusions and perspectives.
    
    \section{Theory}	
    \label{sec:model-problem}
	
    As a reference problem\footnote{Multi-dimensional functions up to second-order tensors will be denoted with bold italic letters.}, the quasi-static elasto-plastic equations (under small deformations) are considered. To this purpose, let us introduce the body $\mathcal{B}$, occupying the spatial region $\Omega \subset \bbR^d$, with $d = 2, 3$, whose boundary is denoted as $\Gamma = \partial \Omega$. Moreover, let $I = (0, T_f)$ be the time interval. The variable to be determined are the displacement field $\vect{u} (\vect{x}, t)$ and the stress field $\ctens{\sigma} (\vect{x}, t)$, with $(\vect{x}, t) \in \Omega \times I$, fulfilling
    \begin{equation}
        \label{eq:model-problem}
        \begin{dcases}
        \div{\ctens{\sigma}} = \vect{f} & \text{in } \Omega \times I \\
        \vect{u} = \vect{u}_D & \text{on } \partial \Omega_D \times I \\
        \ctens{\sigma} \cdot \vect{n} = \vect{f}_N & \text{on } \partial \Omega_N \times I \\
        \vect{u} = \vect{u}_0 & \text{in } \Omega \times \{0\}
        \end{dcases}  
    \end{equation}
    
    where $\vect{f} = \vect{f}(\vect{x}, t)$ is a prescribed history of body forces, $\partial \Omega_D$ and $\partial \Omega_N$ are the Dirichlet and Neumann regions of the boundary $\partial \Omega = \partial \Omega_D \mathbin{\dot{\cup}} \partial \Omega_N$ (where the symbol $\mathbin{\dot{\cup}}$ denotes a disjoint union) and $\vect{n}$ is the outward unit normal vector to $\partial \Omega_N$. As usual, $\vect{u}_0$ is the initial condition, $\vect{u}_D$ is a prescribed displacement on $\partial \Omega_D$ and $\vect{f}_N$ is a prescribed traction (per unit deformed area) on $\partial \Omega_N$.
    
    Moreover, $\ctens{u}$ and $\ctens{\sigma}$ verify the elasto-plastic constitutive relation:
    \begin{equation}
        \label{eq:Hooke}
            \ctens{\sigma} = \ctensd{C} : (\ctens{\varepsilon} - \ctens{\varepsilon}^p)
    \end{equation}
    
    with $\ctensd{C}$ the fourth-order stiffness tensor, $\ctens{\varepsilon} = \nabla^s {{\vect{u}}}$ the total strain tensor ($\nabla^s(\bullet)$ being the symmetric gradient operator), $\ctens{\varepsilon}^p$ the plastic strain tensor and : referring to the tensor product twice contracted.
	
	\subsection{Global equilibrium weak form}
	
    The weak formulation of the quasi-static problem \eqref{eq:model-problem}, under small deformations hypotheses, is easily retrieved from the local form of the equilibrium \cite{plasticity-1,plasticity-2,plasticity-3}.
    
    As first, let us denote with $V_D$ the space of regular enough and kinematically admissible displacement fields (i.e., ensuring the imposed displacement on the region $\partial \Omega_D$):
    \begin{equation}
        V_D = \{ \vect{u} : \Omega \times I \to H^1(\Omega; \bbR^d) \text{ s.t. } \vect{u} (\vect{\gamma}, t) = \vect{u}_D (\vect{\gamma}, t), \quad (\vect{\gamma}, t) \in \partial \Omega_D \times I\}.
    \end{equation}
    
    Similarly, $H^1_0(\Omega; \bbR^d)$ is the space of test functions, satisfying null Dirichlet condition on $\partial \Omega_D$:
    \begin{equation}
        H^1_0(\Omega; \bbR^d) = \{ {\vect{v}} : \Omega \to H^1(\Omega; \bbR^d) \text{ s.t. } {\vect{v}} (\vect{\gamma}) = \vect{0}, \quad \vect{\gamma} \in \partial \Omega_D \}.
    \end{equation}
    
    With these definitions made, the weak formulation seeks $\vect{u} \in V_D$ verifying, $\forall t \in I$ and $\forall {\vect{v}} \in V_0$:
    \begin{equation}
        \label{eq:weak-form}
        \int_\Omega [\ctens{\sigma} (t) : {{\ctens{\varepsilon}}({\vect{v}}) } - \vect{f} (t) \cdot {\vect{v}} ] \dd{{\vect{x}}} - \int_{\partial \Omega_N} \vect{f}_N (t) \cdot {\vect{v}} \dd{{\vect{\gamma}}} = 0.
    \end{equation}
    
    By using equation \eqref{eq:Hooke} and reordering the terms, equation \eqref{eq:weak-form} becomes:
    \begin{equation}
        \label{eq:weak-form-nl}
        \int_{\Omega} \ctens{\varepsilon} ({\vect{v}}) :  \ctensd{C} : \ctens{\varepsilon} (\vect{u}) \dd{{\vect{x}}} - \int_{\Omega} \ctens{\varepsilon} ({\vect{v}}) : \ctensd{C} : \ctens{\varepsilon}^p ({\vect{u}}) \dd{{\vect{x}}} = \int_{\Omega} {\vect{f}}(t) \cdot {\vect{v}} \dd{{\vect{x}}} + \int_{\partial \Omega_N} \vect{f}_N (t) \cdot {\vect{v}} \dd{{\vect{\gamma}}},
    \end{equation}
    or, equivalently,
    \begin{equation}
        \label{eq:weak-form-nl-2}
        k(\vect{u}(t), \vect{v}) - f^{p}(\vect{u}(t), \vect{v}) = f^{ext}(\vect{v}; t),
    \end{equation}
    after having introduced the bilinear and linear forms
    \begin{equation}
    \begin{dcases}
        k(\vect{u}, \vect{v}) = \int_{\Omega} \ctens{\varepsilon} ({\vect{v}}) :  \ctensd{C} : \ctens{\varepsilon} (\vect{u}) \dd{{\vect{x}}} \\
        f^{ext}(\vect{v}; t) = \int_{\Omega} \vect{f}(t) \cdot \vect{v} \dd{{\vect{x}}} + \int_{\partial \Omega_N} \vect{f}_N (t) \cdot {\vect{v}} \dd{{\vect{\gamma}}},
    \end{dcases}
    \end{equation}
    as well as the nonlinear term accounting for the plastic strain
    \begin{equation}
        f^{p}(\vect{u}, \vect{v}) = \int_{\Omega} \ctens{\varepsilon} ({\vect{v}}) : \ctensd{C} : \ctens{\varepsilon}^p (\vect{u}) \dd{{\vect{x}}}.
    \end{equation}
	
	Let us recall that evaluation of $f^{p}(\vect{u}(t), \vect{v})$ in equation \eqref{eq:weak-form-nl-2} requires the knowledge of $\ctens{\varepsilon}^p (\vect{u}(t))$, which depends on the assumed plasticity model \cite{plasticity-1,plasticity-2,plasticity-3}.
		
	\subsection{Plasticity model}
	
	Here a standard von Mises ($J_2$) plasticity, whose yield surface is defined by
	\begin{equation}
		\label{eq:yield_surface}
		\Phi(\ctens{\sigma}(t), \ctens{\varepsilon}^p(t) ) = \sqrt{3 J_2} - \sigma_{y, t} = 0,
	\end{equation}
	where $J_2 = J_2(t)$ denotes the second deviatoric invariant 
	\begin{equation}
		J_2 = J_2(\ctens{s}) = \frac{1}{2} \ctens{s}:\SDtens{s}, \quad \ctens{s} = \ctens{\sigma} - \frac{1}{3} \tr(\ctens{\sigma}) \ctens{I}
	\end{equation}
	and $\sigma_{y, t}$ is the uniaxial yield stress, which evolves through a suitable strain-hardening curve. The details of this evaluation are given below.
	
	Making the assumption of isotropic hardening, at any state of hardening, the evolution of the yield surface \eqref{eq:yield_surface} corresponds to a uniform (isotropic) expansion of the initial one. This is obtained assuming $\sigma_{y, t}$ being a function of the accumulated (or effective) plastic strain
	\begin{equation}
		\label{eq:accumulated_plastic}
		\bar{\varepsilon}^p_t = \int_0^t \sqrt{\frac{2}{3} \dot{\ctens{\varepsilon}}^p : \dot{\ctens{\varepsilon}}^p } \dd{s}.
	\end{equation}
	In particular, assuming linear hardening, $\sigma_{y, t}$ is given by 
	\begin{equation}
		\label{eq:linear_hardening}
		\sigma_{y, t} = \sigma_{y, 0} + H \bar{\varepsilon}^p_t,
	\end{equation}
	where $\sigma_{y, 0}$ is the initial yield stress and $H$ is the (constant) hardening modulus.
	
	Additionally, a standard associative plastic flow rule is considered, meaning that the plastic strain rate is a tensor normal to the yield surface in the stress space, that is
	\begin{equation}
		\label{eq:plastic_multiplier}
		\dot{\ctens{\varepsilon}}^p = \dot{\lambda} \ctens{N},\quad \ctens{N} \defeq \pdv{\Phi}{\ctens{\sigma}} = \sqrt{\frac{3}{2}} \frac{\ctens{s}}{\norm{\ctens{s}}}.
	\end{equation}
	Following usual notations, in equation \eqref{eq:plastic_multiplier}, $\dot{\lambda}$ denotes the unknown plastic multiplier \cite{plasticity-2}. 

    \section{Material and methods}
    \label{sec:PGD-based-time-multiscale}
    
    In most finite element based approaches, the equilibrium \eqref{eq:weak-form-nl-2} is restored incrementally. This corresponds to a global loop, where the nonlinearity due to $f^p$ is tackled either explicitly or implicitly \cite{plasticity-1, plasticity-2}. However, nonlinear problems such as \eqref{eq:weak-form-nl-2} have also been successfully addressed using non-incremental strategies in time, as suggested by the LATIN and PGD literature \cite{pgd-multiscale-2, PGD-1, PGD-elastoplastic, PGD-elastoplastic-1, PGD-nonlinear-dynamics}. Here, the solution is expressed in the low-rank separated form 
    \begin{equation}
    	\label{eq:space-time-solution}
    	\vect{u}_m(\vect{x}, t) = \sum_{k = 1}^m \vect{U}_k^\vect{x}(\vect{x}) U_k^t(t),
    \end{equation}
	and computed directly in the whole space-time domain, by means of an iterative strategy. In equation \eqref{eq:space-time-solution}, $m$ denotes the rank of the solution, also known as number of PGD modes. 
    
    In this work, the same linearization strategy of PGD approaches is employed and a further time multiscale representation of the solution is addressed.
    
    As usual in computational plasticity, the integration of the constitutive equations corresponds to a local loop, usually referred as state-updating procedure \cite{plasticity-1, plasticity-2}. Indeed, the plasticity model reduces to differential constitutive equations which may be solved numerically by means of an Euler scheme (explicit or implicit). Here, an implicit algorithm based on the elastic predictor/return-mapping procedure is adopted, where the resulting nonlinear equation (for the incremental plastic multiplier) is tackled by a Newton-Raphson scheme.
    
    \subsection{Linearization and space-time separation}
    \label{subsec:nonlinearity-space-time}
    
    Following a standard Galerkin approach, the approximation of $\vect{u}(t)$ satisfying \eqref{eq:weak-form-nl-2} is sought in the finite dimensional subspace of $V_D$ defined as
    \begin{equation}
    	V_D^h = \left\{ \vect{u}_h \in V_D \text{ s.t. } \vect{u}_h (\vect{x}, t) = \sum_{i = 1}^{N_{\vect{x}}} \vect{u}_i(t) \phi_i (\vect{x}), \quad (\vect{x}, t) \in \Omega \times I  \right\}
    \end{equation}
	where $\{\phi_i\}_{i=1}^{N_{\vect{x}}}$ is the set of suitably chosen shape functions in space and $\{\vect{u}_i(t)\}_{i=1}^{N_{\vect{x}}}$ the corresponding unknown temporal coefficients. Similarly, $V_0^h$ denotes the Galerkin approximation space of $H^1_0(\Omega; \bbR^d)$.
	
	Within the semi-discrete counterpart of \eqref{eq:weak-form-nl-2}, one looks for $\vect{u}_h \in V_D^h$ such that, $\forall t \in I$ and $\forall \vect{v}_h \in V_0^h$,
	\begin{equation}
		\label{eq:weak-form-nl-2-semidiscrete}
		k(\vect{u}_h(t), \vect{v}_h) - f^{p}(\vect{u}_h(t), \vect{v}_h) = f^{ext}(\vect{v}_h; t).
	\end{equation}
	
    To tackle the nonlinearity of the problem, the first step consists in computing an approximation of the elastic solution verifying\footnote{For the sake of notational simplicity, the subscript $h$ related to the Galerkin approximation is suppressed.}, $\forall t \in I$
    \begin{equation}
        \label{eq:weak-form-elastic}
        k(\vect{u}^{(0)}(t), \vect{v}_h) = f^{ext} (\vect{v}_h; t),
    \end{equation}
	whose algebraic counterpart is straightforwardly obtained as
    \begin{equation}
    \label{eq:algebraic-nonlinear}
        \SDvect{K} \SDvect{u}^{(0)}(t) = \SDvect{f}^{ext}(t),
    \end{equation}
    where $K_{ij} = k(\phi_i, \phi_j)$ and $f_{i}^{ext}(t) = f^{ext}(\phi_i; t)$, for $i, j = 1, \dots, N_{\vect{x}}$.
    
    At this point, a temporal discretization of the interval $I$ is introduced. This is based on considering $N_t$ uniform times $t_i$, such that $t_{i+1} - t_i = \Delta t > 0$, for $i = 1, \dots, N_t$.

    Equation \eqref{eq:algebraic-nonlinear} can then be rewritten in a tensorial formalism over the whole space-time domain \cite{pgd-multiscale}, looking for $\SDvect{U}^{(0)} \in \bbR^{N_{\vect{x}} \times N_t}$ such that
    \begin{equation}  
        \label{eq:algebraic-nonlinear-s-t}
        ( \SDvect{K} \otimes \SDvect{I}_{N_t} ) : \SDvect{U}^{(0)} = \SDvect{F}^{ext}
    \end{equation}
    where $\SDvect{I}_{N_t} \in \bbR^{N_t \times N_t}$ denotes the identity matrix in time and $\SDvect{F}^{ext} \in \bbR^{N_{\vect{x}} \times N_t}$ collects the time evaluations of the right-hand-side in \eqref{eq:algebraic-nonlinear}, that is $\SDvect{f}^{ext}(t_i) \in \bbR^{N_{\vect{x}}}$, $i = 1, \dots, N_t$. 

   Problem \eqref{eq:algebraic-nonlinear-s-t} can be treated directly by means of the PGD algorithm, which seeks a low-rank separated approximation of $\SDvect{U}^{(0)}$ as
    \begin{equation}
    	\label{eq:pgd-approx-s-t}
        \SDvect{U}^{(0)} \approx \SDvect{U}^{(0)}_{m^{(0)}} = \sum_{k = 1}^{m^{(0)}} \SDvect{U}^{(0), \vect{x}}_k \otimes \SDvect{U}^{(0), t}_k,
    \end{equation}
	where $m^{(0)}$ is the number of PGD modes. For the PGD assembly and solution of space-time separated problem, the reader may refer to \cite{pgd-multiscale} and references therein.

    Since $\SDvect{U}^{(0)}_{m^{(0)}}$ is a low-rank approximation of $\vect{u}^{(0)} (t)$, $\forall t \in I$. An approximation of the elastic solution $\vect{u}^{(0)} (t)$, $\forall t \in I$ all history-dependent variables may be updated from $\ctens{\varepsilon}^{(0)}= \nabla^s \vect{u}^{(0)}$. In particular, the update of $\ctens{\varepsilon}^{p, (0)}$ allows to freeze the nonlinear term in equation \eqref{eq:weak-form-nl-2} and start an iterative process where each iteration $l \geq 1$ consists of two steps:
    \begin{enumerate}
        \item (\textit{state-updating}) The update of $\ctens{\varepsilon}^{p, (l - 1)} = \ctens{\varepsilon}^{p}(\vect{u}^{(l-1)})$ via the elastic predictor/return-mapping procedure. This consists of a set of nonlinear evaluations which can be written as
        \begin{equation}
        	\ctens{\varepsilon}^{p, (l - 1)} =  \mathcal{N}(\ctens{\varepsilon}^{(l - 1)}, \bar{\varepsilon}^p_{T_f}),
        \end{equation} 
    	where $\mathcal{N}$ represents the nonlinear operator depending on the total strain tensor and on the effective plastic strain up to the final time $T_f$.  	
        
        \item (\textit{linearized equilibrium}) The solution of the linearized problem
        \begin{equation}
            \label{eq:weak-form-linearized}
                k(\vect{u}^{(l)}(t), \vect{v}_h) = f^{ext}(\vect{v}_h; t) + f^{p}(\vect{u}^{(l-1)}(t), \vect{v}_h),
            \end{equation}
        looking for the space-time separated approximation
        \begin{equation}
        	\label{eq:space-time-algebraic}
        	\SDvect{U}^{(l)} \approx \SDvect{U}^{(l)}_{m^{(l)}} = \sum_{k = 1}^{m^{(l)}} \SDvect{U}^{(l), \vect{x}}_k \otimes \SDvect{U}^{(l), t}_k
        \end{equation}
    	of the corresponding tensorial problem
    	\begin{equation}  
    		\label{eq:algebraic-nonlinear-s-t-2}
    		( \SDvect{K} \otimes \SDvect{I}_{N_t} ) : \SDvect{U}^{(l)} = \SDvect{F}^{ext} + \SDvect{F}^{p, (l - 1)}
    	\end{equation}
    	with $\SDvect{F}^{p, (l - 1)} \in \bbR^{N_x \times N_t}$ accounting for the plastic contributions. 
    \end{enumerate}
	The linearization loop stops when two successive approximations become close enough under a suitable distance. This means that, given a small enough $\delta > 0$
	\begin{equation}
		\label{eq:convergence-criterion}
		e_{l} = \frac{\norm{ \SDvect{U}^{(l)} - \SDvect{U}^{(l - 1)}}_F}{\norm{ \SDvect{U}^{(l - 1)} }_F} < \delta,
	\end{equation} 
	where $\norm{\bullet}_F$ denotes, for instance, the standard Frobenius norm.
	
	Denoting with $L$ the iteration satisfying the convergence criterion \eqref{eq:convergence-criterion}, the final PGD approximation of the nonlinear problem \eqref{eq:weak-form-nl-2} is $\SDvect{U}^{(L)}$, that is the approximation \eqref{eq:space-time-solution} has rank $m = m^{(L)}$.
	
\begin{figure}
		\centering
		\includegraphics[width=0.5\textwidth]{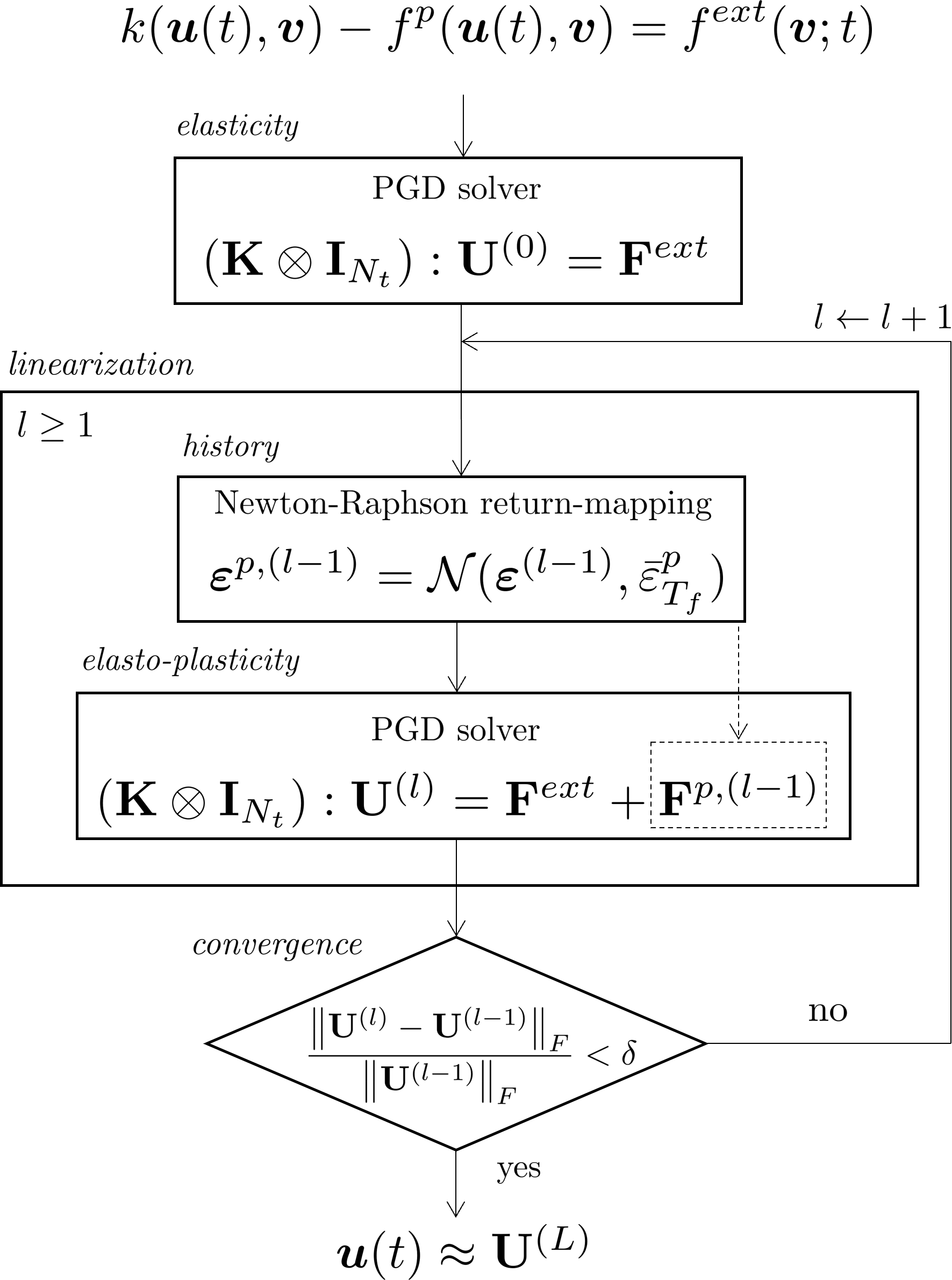}
		\caption{PGD-based solving scheme for elasto-plasticity.}
		\label{fig:solving-scheme}
	\end{figure}

	The overall solving procedure is summarized in the flowchart in figure \ref{fig:solving-scheme}.
	
	\begin{remark}
		The assembly of the stiffness matrix $\SDvect{K}$ is a classic finite element based one and the evaluation of right-hand-side $\SDvect{F}^{p, (l - 1)}$ follows a standard plasticity integration algorithm (elastic predictor/return-mapping procedure). Both these computations can be performed using any computational mechanics software. The overall procedure is thus weakly-intrusive, since it only requires the externalization of the linearization loop and the usage of the space-time PGD solver after the assembly.
	\end{remark}

	\begin{remark}
		As discussed in \cite{pgd-multiscale}, a further decomposition of the time axis can be easily enforced in the PGD constructor. This becomes particularly interesting when the system response is expected to exhibit multiple time scales. Moreover, from a computational viewpoint, the solution of the linearized problem benefits of operational and memory savings, especially when combining large time intervals with small time steps.
	\end{remark}

    With the aim of reducing the overall complexity of the solving scheme, the next subsection investigates the existance of a time multiscale decomposition of the elastoplastic response.

    \subsection{Multi-time separation}

     Following the same strategy of \cite{pgd-multiscale, pgd-multiscale-1, pgd-multiscale-2}, the computation of the PGD time modes $\{U_i^t(t)\}_{i=1}^{m}$ can be addresssed via a multi-time separated representation (MT-PGD).

    In particular, the PGD solution of the linearized problem (the index of the nonlinear iteration is suppressed for notational simplicity) may be approximated via a space-microtime-macrotime separated representation \cite{pgd-multiscale, pgd-multiscale-1}
    \begin{equation}
        \label{eq:space-multimodes-triple}
        \vect{u}_m(\vect{x}, t) = \sum_{k = 1}^m \vect{U}^{\vect{x}}_k(\vect{x}) U^t_k(t) \approx \sum_{k = 1}^M \vect{U}^{\vect{x}}_k(\vect{x}) U^\tau_k (\tau) U^T_k(T),
    \end{equation}
    or imposing the multi-time decomposition for the computation of the time function (micro/macro time sub-modes) \cite{pgd-multiscale-1, pgd-multiscale-2}
    \begin{equation}
        \label{eq:space-multimodes}
        \vect{u}_m(\vect{x}, t) = \sum_{k = 1}^m \vect{U}^{\vect{x}}_k(\vect{x}) U^t_k(t) \approx \sum_{k = 1}^m \vect{U}^{\vect{x}}_k(\vect{x}) \sum_{j = 1}^{m_k} U^\tau_{k, j}(\tau) U^T_{k, j}(T),
    \end{equation}
    where two new independent time coordinates $\tau$ (microtime, or fast time) and $T$ (macrotime, or slow time) have been introduced. Moreover, in equation \eqref{eq:space-multimodes-triple} the new number of modes is denoted as $M$ (in general, $M > m$), while in equation \eqref{eq:space-multimodes} the number of modes involved in the space-time separation is still $m$ and $m_k$ denotes the number of time sub-modes required to approximate the single-scale function for the current global mode $k$. 
    
    At the discrete level, the usual temporal discretization $\{t_i\}_{i = 0}^{N_t}$ of the interval $I$ is now replaced by
    \begin{enumerate}
        \item a macro-discretization $\{T_i\}_{i = 0}^{N_T}$ of equispaced macro-dofs spanning the whole interval $I$, with a constant macro-step $\Delta T = T_{i + 1} - T_i > 0$, $T_0 = 0$;
        \item a micro-discretization $\{\tau_i\}_{i = 0}^{N_{\tau}}$ of equispaced micro-dofs spanning the first macro subinterval $[0, T_1)$, with a constant micro-step $\Delta \tau = \tau_{i + 1} - \tau_i > 0$, $\tau_0 = T_0 = 0$ and $\tau_{N_t} = T_1 - \Delta \tau$.
    \end{enumerate}
    In such a way, the single-scale fine time grid is recovered by the tensor product of the two newly introduced micro and macro grids, that is $\{t_i\}_{i = 0}^{N_t} = \{T_i\}_{i = 0}^{N_T} \otimes \{\tau_i\}_{i = 0}^{N_{\tau}}$ meaning $N_t = N_T N_{\tau}$.
    
    \begin{remark}
    	Within the PGD assembly, when considering time multiscale approximations, the time operator $\mathbf{I}_t$ is exactly recovered as $\mathbf{I}_{N_t} = \mathbf{I}_{N_T} \otimes \mathbf{I}_{N_\tau}$. Moreover, when considering PDEs involving time derivatives, exact tensorial decompositions of the related time operators have been discussed in \cite{pgd-multiscale}.
    \end{remark}
	
	\begin{remark}
		In approximation \eqref{eq:space-multimodes}, each product $U^\tau_{k, j}(\tau) U^T_{k, j}(T)$ is not granted to be continuous by construction. The tensor product is, indeed, replicating the microscale patterns $U^\tau_{k, j}(\tau)$ along the macroscale. However, the continuity of the multi-time approximation of $U^t_k(t)$ is ensured by adding a sufficient number of modes $m_k$.
	\end{remark}
    
    \begin{remark}
    Both approaches \eqref{eq:space-multimodes-triple} and \eqref{eq:space-multimodes} guarantee reduced storage requirements since $N_t$ function evaluations are reconstructed combining $N_T$ and $N_\tau$ evaluations of the macro and micro functions. Moreover, as usual in PGD methods, the size of independent systems to be solved in the alternating direction strategy (ADS) is now smaller \cite{pgd-multiscale}, implying computational savings. 
    \end{remark}

    \begin{remark}
    If the single-scale function $U_k^t(t)$ exhibits a multiscale behaviour and a wise choice of the macro-partitions is done (e.g., respecting typical patterns of the function such as the ones coming from a periodic behaviour), only a few sub-modes $m_k$ are enough to have a good approximation \eqref{eq:space-multimodes} \cite{pgd-multiscale-2}.
    \end{remark}
    
    Using representation \eqref{eq:space-multimodes} in this case is sufficient because the primary goal of this work is to investigate multiscale patterns inside PGD time functions. In this PGD formalism, it shall be noticed that a discrete function of time $\SDvect{h}^t \in \bbR^{N_t}$ is now expressed as 
    \begin{equation}
        \SDvect{h}^t \approx \sum_{j = 1}^{m_k} \SDvect{h}^{T} \otimes \SDvect{h}^{\tau},
    \end{equation}
    guaranteeing important computational savings since the operations to be performed along the new separated coordinates involve lower number of dofs. 

    Particularly, the procedure presented in subsection \ref{subsec:nonlinearity-space-time} is extended to a time multiscale framework by replacing the approximation \eqref{eq:space-time-algebraic} with its algebraic multi-time counterpart
    \begin{equation}
        \label{eq:space-multitime-algebraic}
        \SDvect{U}^{(l)} \approx \SDvect{U}^{(l)}_{m^{(l)}} = \sum_{k = 1}^{m^{(l)}} \SDvect{U}^{(l), \vect{x}}_i \otimes \sum_{j = 1}^{m_k} \SDvect{U}^{(l), T}_{k, j} \otimes \SDvect{U}^{(l), \tau}_{k, j}.
    \end{equation}

    This straightforwardly entails a reduced time multiscale representation of the total strain
    \begin{equation}
        \label{eq:eps-space-multimodes}
        \ctens{\varepsilon}_m(\vect{x}, t) = \sum_{k = 1}^m \ctens{\varepsilon}^{\vect{x}}_k(\vect{x}) \varepsilon^{t}_k(t) \approx \sum_{k = 1}^m \ctens{\varepsilon}^{\vect{x}}_k(\vect{x}) \sum_{j = 1}^{m_k} \varepsilon^{\tau}_{k, j}(\tau) \varepsilon^{T}_{k, j}(T).
    \end{equation}
	
	\begin{remark}
		It shall be noticed that, so far, the computational gains entailed by the usage of the MT-PGD are the ones discussed in \cite{pgd-multiscale, pgd-multiscale-2} for linear problems. Indeed, the nonlinearity the step 1 (state-updating) requires a standard integration over $N_t$ steps because of the history-dependency. In other terms, so far, the MT-PGD is not exploited for the evaluation of the elastoplastic constitutive relation. In the workflow in figure \ref{fig:solving-scheme}, the PGD solver blocks are replaced by the MT-PGD ones.
		
		Our works in progress are targeting the computational effectiveness from a original evaluation of the nonlinear term.
	\end{remark}
    
	
    \section{Results and discussion}
    \label{sec:application-case}
    
    In this section, two examples in cyclic elasto-plasticity are considered, varying the geometry of the specimen and the imposed loading. The problems are solved numerically via the PGD space-time constructor. Special care is paid to the PGD time modes and their characterization by means of a multi-time decomposition (MT-PGD). 
    
    \subsection{Dog-bone shaped specimen}

    A uniaxial load-unload tensile test over a dog-bone shaped steel specimen under monoperiodic cyclic loading is here considered. The loading consists in a Dirichlet datum $u_D(t)$ having constant amplitude applied to both sides of the specimen. 
   	
   	The material has a Young modulus $E = 210$ GPa and a Poisson ratio $\nu = 0.3$. The assumed plasticity law \eqref{eq:linear_hardening} is characterized by an initial yield stress $\sigma_{y, 0} = 205$ MPa and a linear hardening coefficient $H = 2$ GPa. To ensure small deformations, the imposed constant amplitude displacement has a maximum amplitude $u_D^{max} = 0.125$ mm and the load rate is fixed at $v_l = 0.025$ mm/s following the standard of quasi-static testing. A single cycle (load-unload-load) time has duration $T_1 = 4 u_D^{max}/v_l = 20$ s.
   
\begin{figure}[H]
	\centering
	\begin{subfigure}{0.45\textwidth}
		\centering
		\includegraphics[width=1\textwidth]{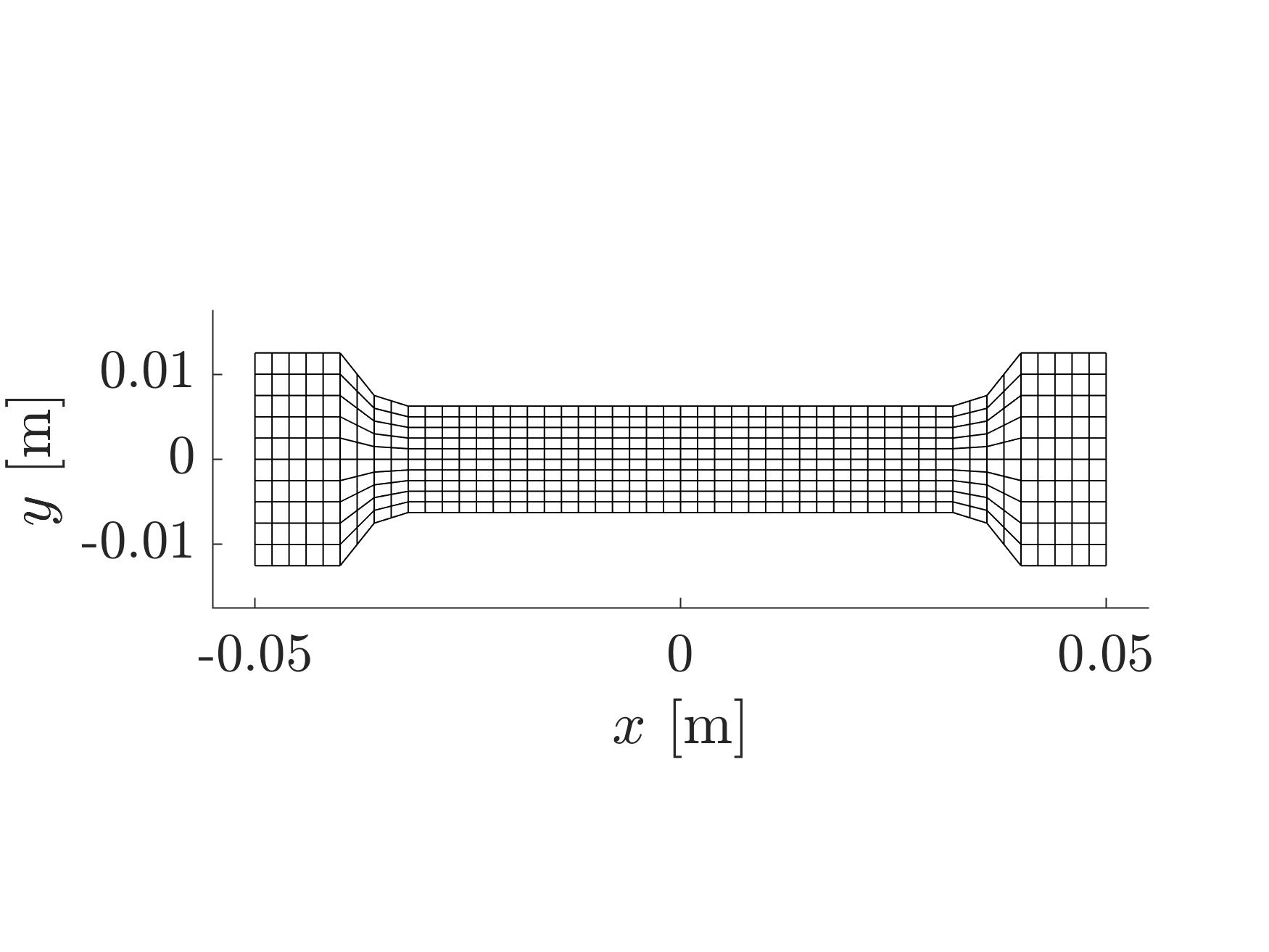}
	\end{subfigure}
	\begin{subfigure}{0.45\textwidth}
		\centering
		\includegraphics[width=1\textwidth]{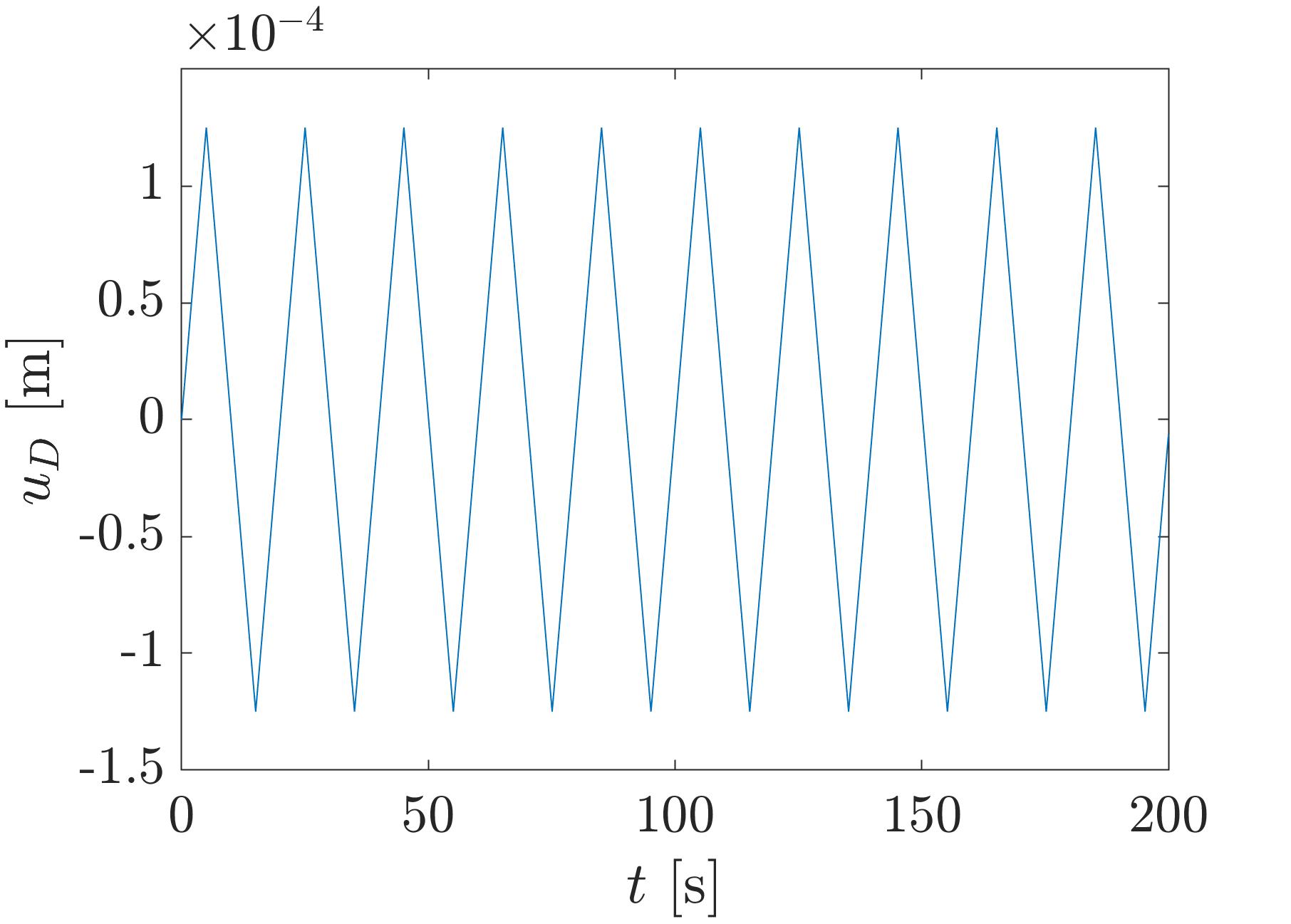}
	\end{subfigure}
	\caption{Discretized geometry (left) and imposed displacement (right).}
	\label{fig:test-case}
    \end{figure}

   	Figure \ref{fig:test-case} shows the two-dimensional discretized geometry and the imposed displacement having 10 cycles. The spatial mesh consists of $N_e = 500$ quadrilateral elements and $N_x = 561$ mesh nodes. The time interval is divided in $N_t = 800$ times.

\begin{figure}[H]
		\centering
		\begin{subfigure}{0.45\textwidth}
			\centering
			\includegraphics[trim={0 300 0 300},clip,width=1\textwidth]{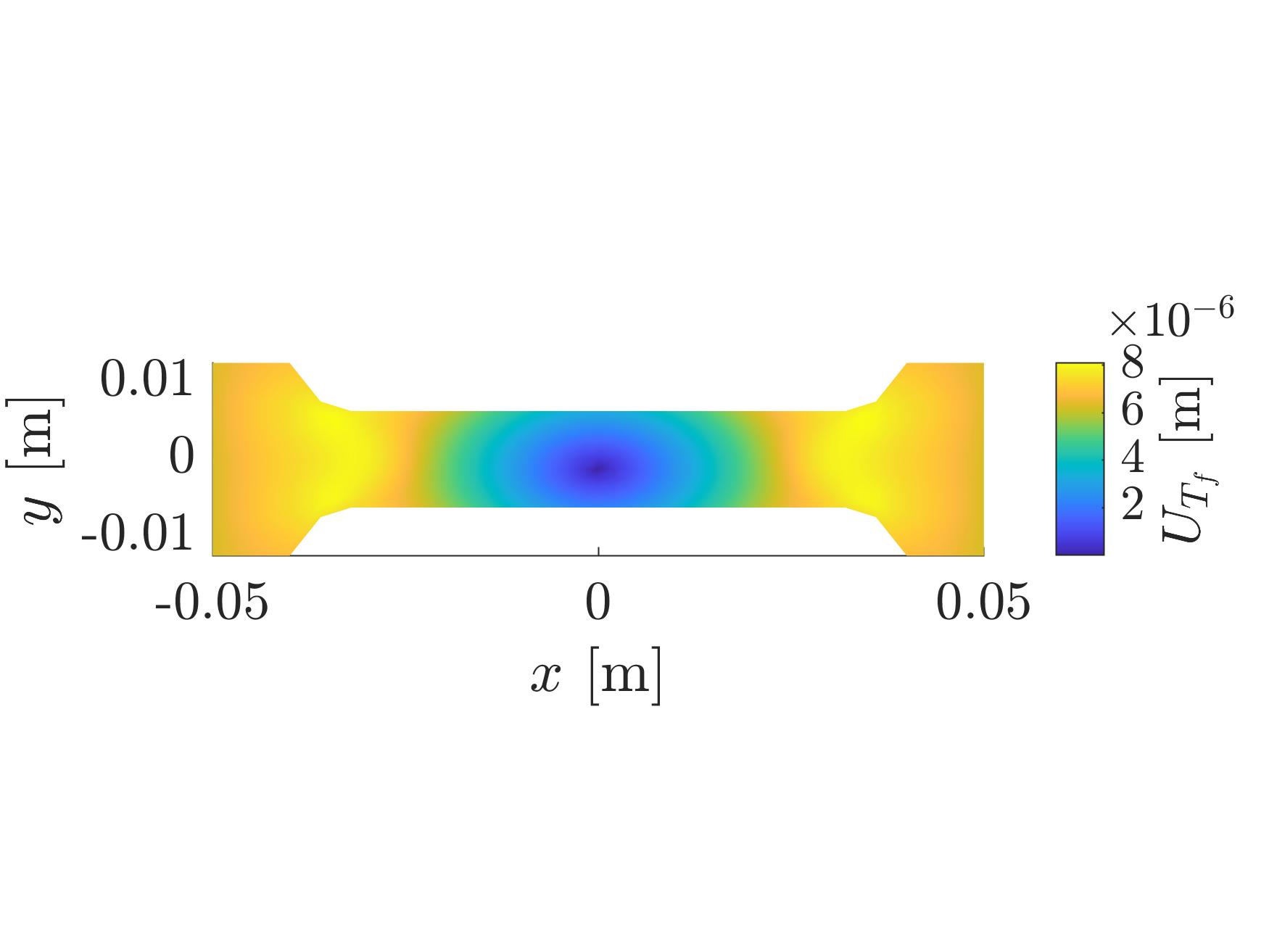}
		\end{subfigure}
		\begin{subfigure}{0.45\textwidth}
			\centering
			\includegraphics[trim={0 300 0 300},clip,width=1\textwidth]{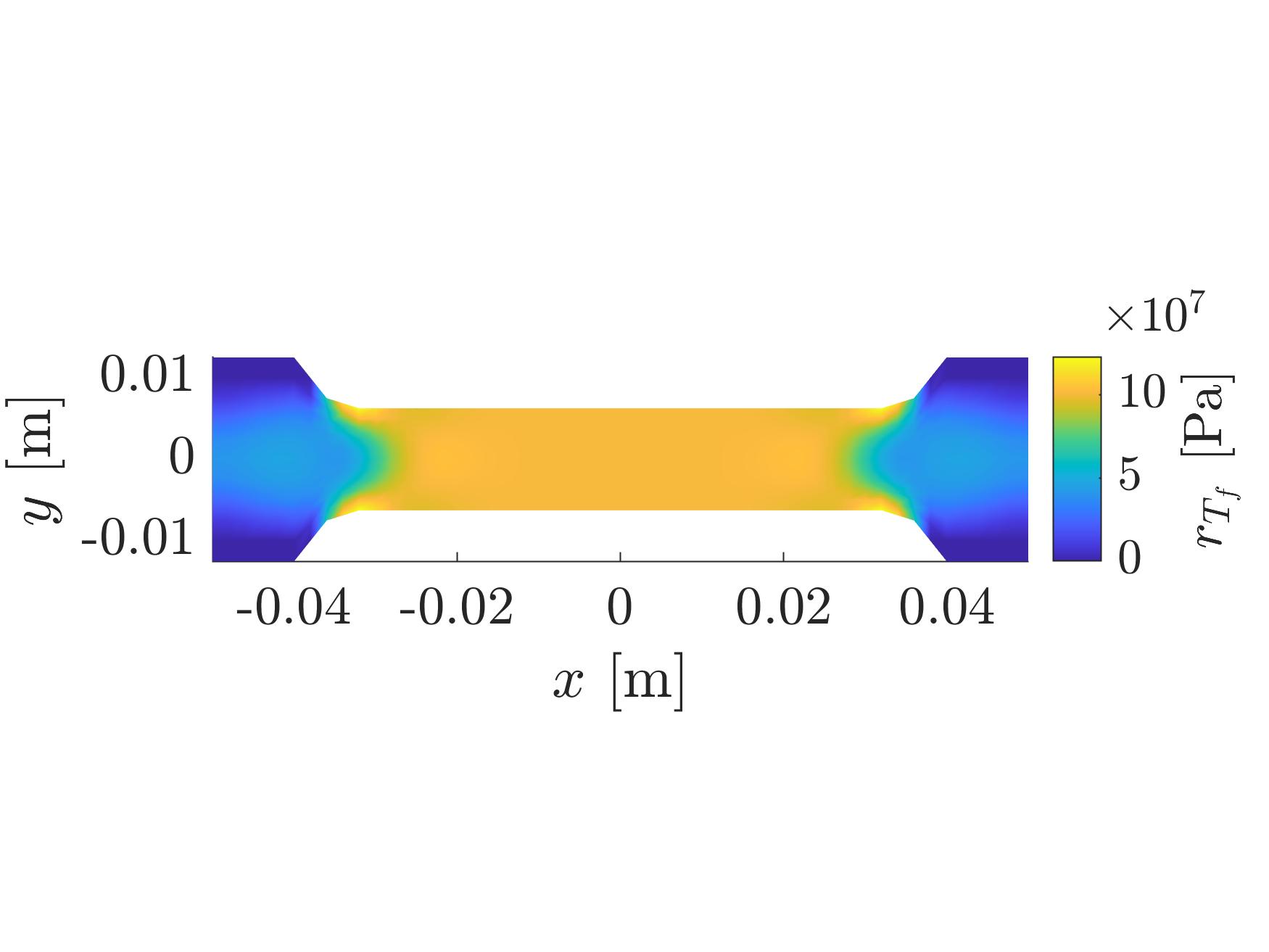}
		\end{subfigure}
		\caption{Displacement field (left) and isotropic hardening (right) at final time $T_f$.}
		\label{fig:test-case-results}
	\end{figure}

	Figure \ref{fig:test-case-results} gives the magnitude of the displacement field and the isotropic hardening function computed at the final time $T_f = 200$ s.

\begin{figure}[H]
		\centering
		\begin{subfigure}{0.45\textwidth}
			\centering
			\includegraphics[width=1\textwidth]{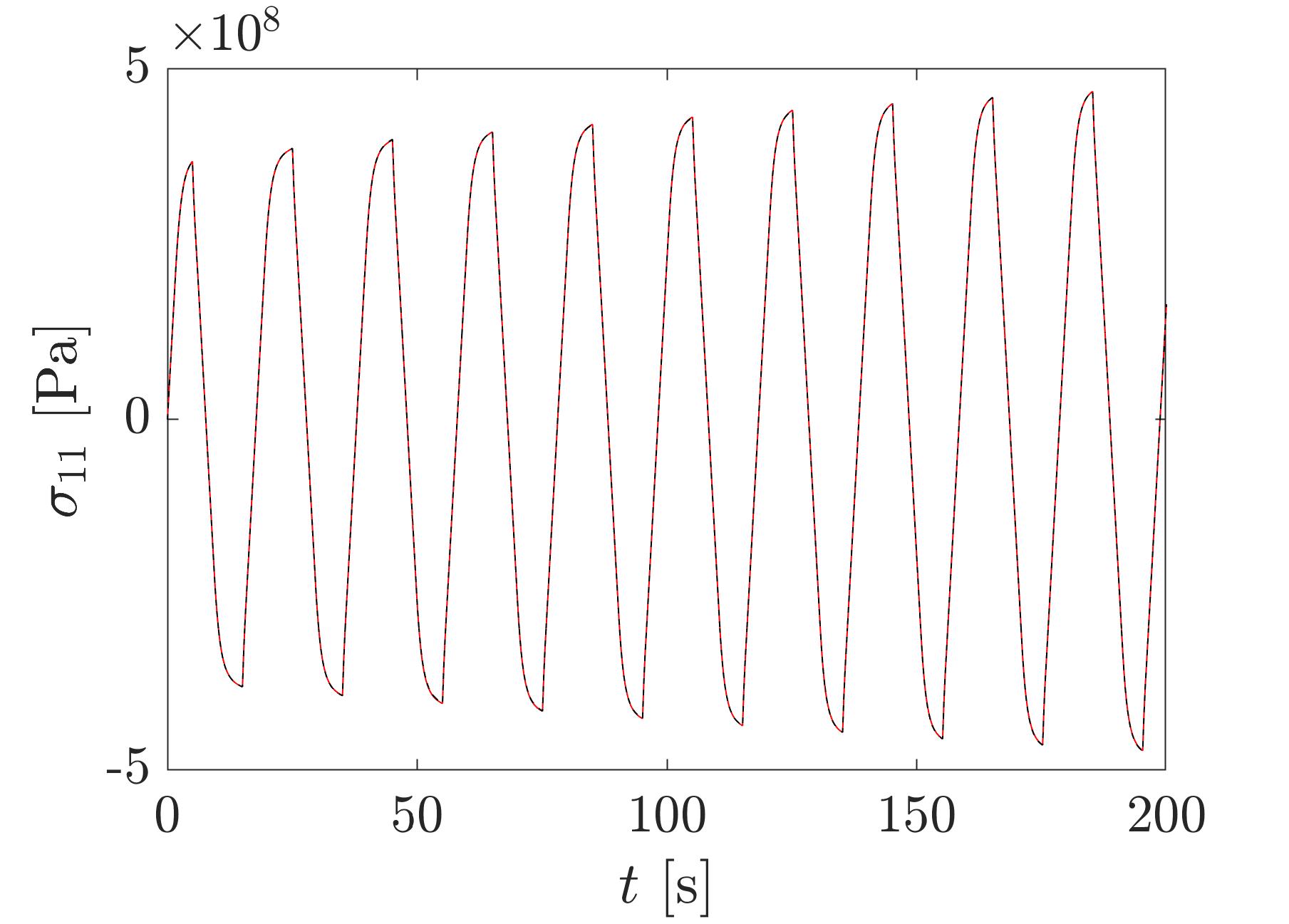}
		\end{subfigure}
		\begin{subfigure}{0.45\textwidth}
			\centering
			\includegraphics[width=1\textwidth]{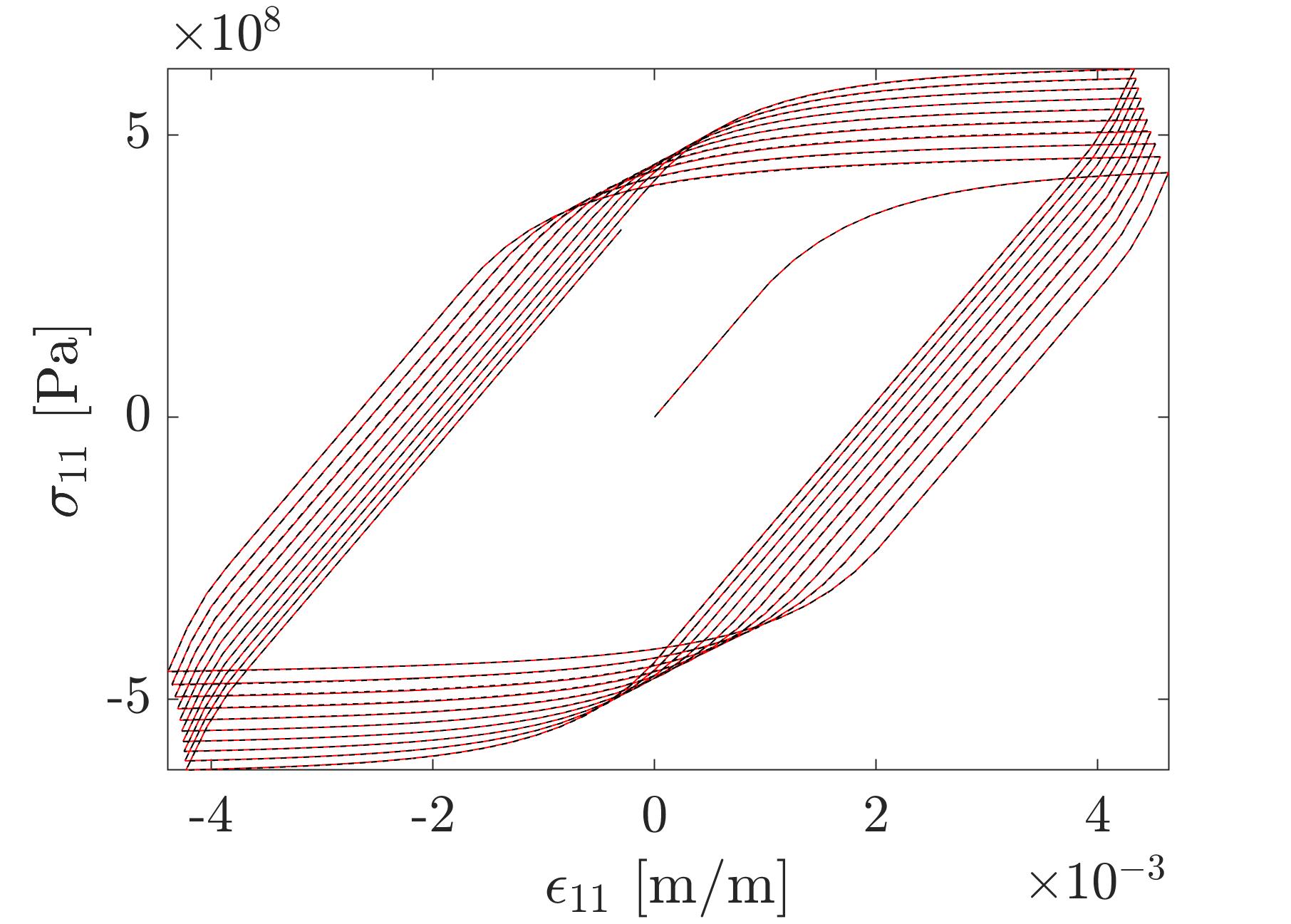}
		\end{subfigure}
		\caption{Stress-displacement curve (left) and hysteresis loop (right) in $(0,0)$. Red line: FE, black dashed line: PGD.}
		\label{fig:local-response}
	\end{figure}

	Figure \ref{fig:local-response} shows the comparison of the PGD results with a classical FE-based incremental algorithm (considering implicit integration of plasticity based on return-mapping) computed at the center of the specimen $(x_0, y_0) = (0,0)$.

\begin{figure}[H]
		\centering
		\includegraphics[width=1\textwidth]{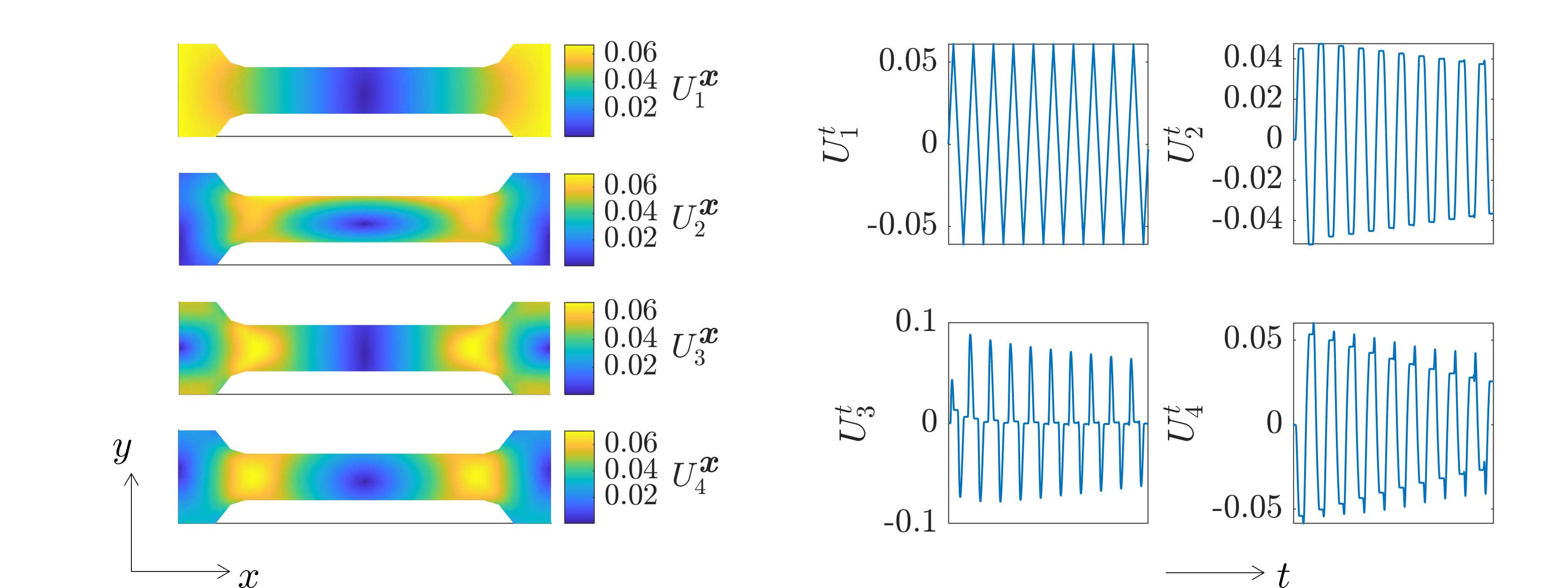}
		\caption{First four normalized PGD modes.}
		\label{fig:pgd-modes}
	\end{figure}

	The first four normalized modes in space and time of the PGD approximation \eqref{eq:space-time-solution} are shown in figure \ref{fig:pgd-modes}. 
	
	Focusing on the time modes, one can observe that $U^t_1$ corresponds to the elastic response, while subsequent modes exhibit similar evolution, characterized by (a) a transient zone at the beginning, (b) a pattern stabilization towards an almost-periodic (cyclic) behavior, (c) a slow decay of the signals amplitude. This behaviour can be easily characterized by the MT-PGD approximation \eqref{eq:space-multimodes}: 
	\begin{equation}
		\vect{u}_m(\vect{x}, t) = \sum_{k = 1}^m \vect{U}^{\vect{x}}_k(\vect{x}) U^t_k(t) \approx \sum_{k = 1}^m \vect{U}^{\vect{x}}_k(\vect{x}) \sum_{j = 1}^{m_k} U^\tau_{k, j}(\tau) U^T_{k, j}(T).
	\end{equation}
	
\begin{figure}[H]
		\centering
		\includegraphics[width=0.5\textwidth]{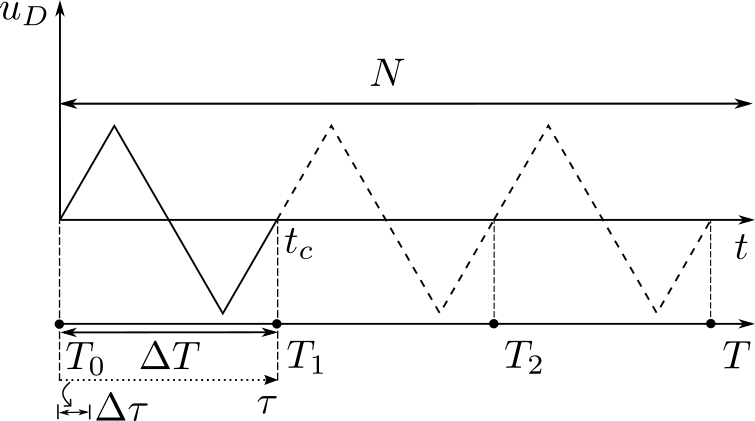}
		\caption{Microscale and macroscale time discretization.}
		\label{fig:micro-macro-mesh}
	\end{figure}

	This is achieved through a macro-discretization based on $N_T$ equispaced macro-times, while the micro-scale consists of $N_\tau$ equispaced micro-times. For instance, if a single cycle is defined from the sequence loading-unloading-loading, the macro-discretization could consists (as later commented out, without losing generality) of a coarse mesh having a macro timestep $\Delta T$ covering a whole cycle. The micro-discretization corresponds to a fine mesh along the cycle, as illustrated in figure \ref{fig:micro-macro-mesh}.
	
\begin{figure}[H]
		\centering
		\includegraphics[width=0.5\textwidth]{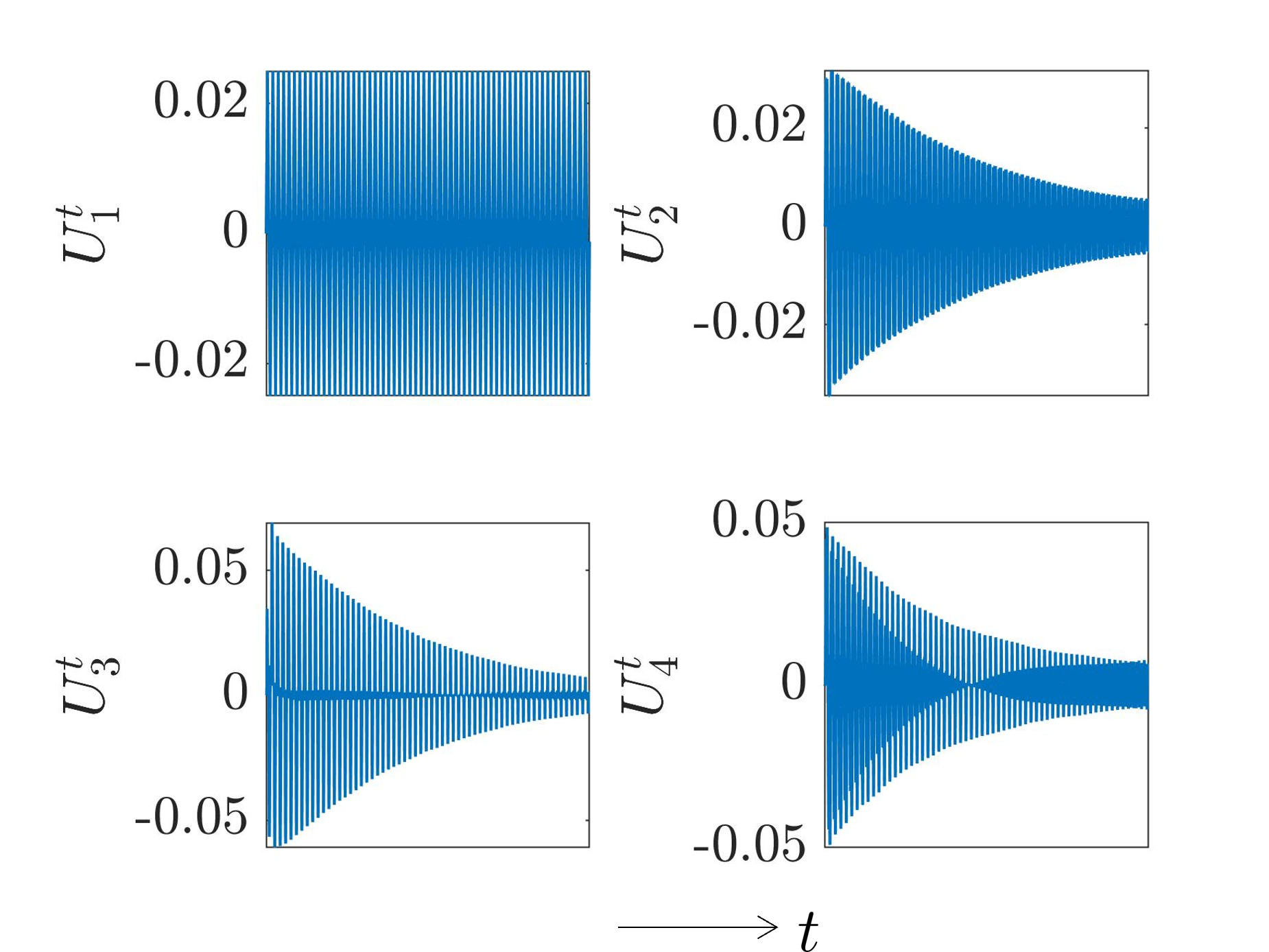}
		\caption{PGD time modes $\{U^t_k(t)\}_{k = 1}^4$ with 60 cycles.}
		\label{fig:time-modes-mtpgd-60}
	\end{figure}

	The advantage of considering such a decomposition is evident when increasing the number of cycles. For instance, figure \ref{fig:time-modes-mtpgd-60} is the counterpart of the time modes figure \ref{fig:pgd-modes} when imposing the same loading over 60 cycles.
	
\begin{figure}[H]
		\centering
		\begin{subfigure}{0.45\textwidth}
			\centering
			\includegraphics[width=1\textwidth]{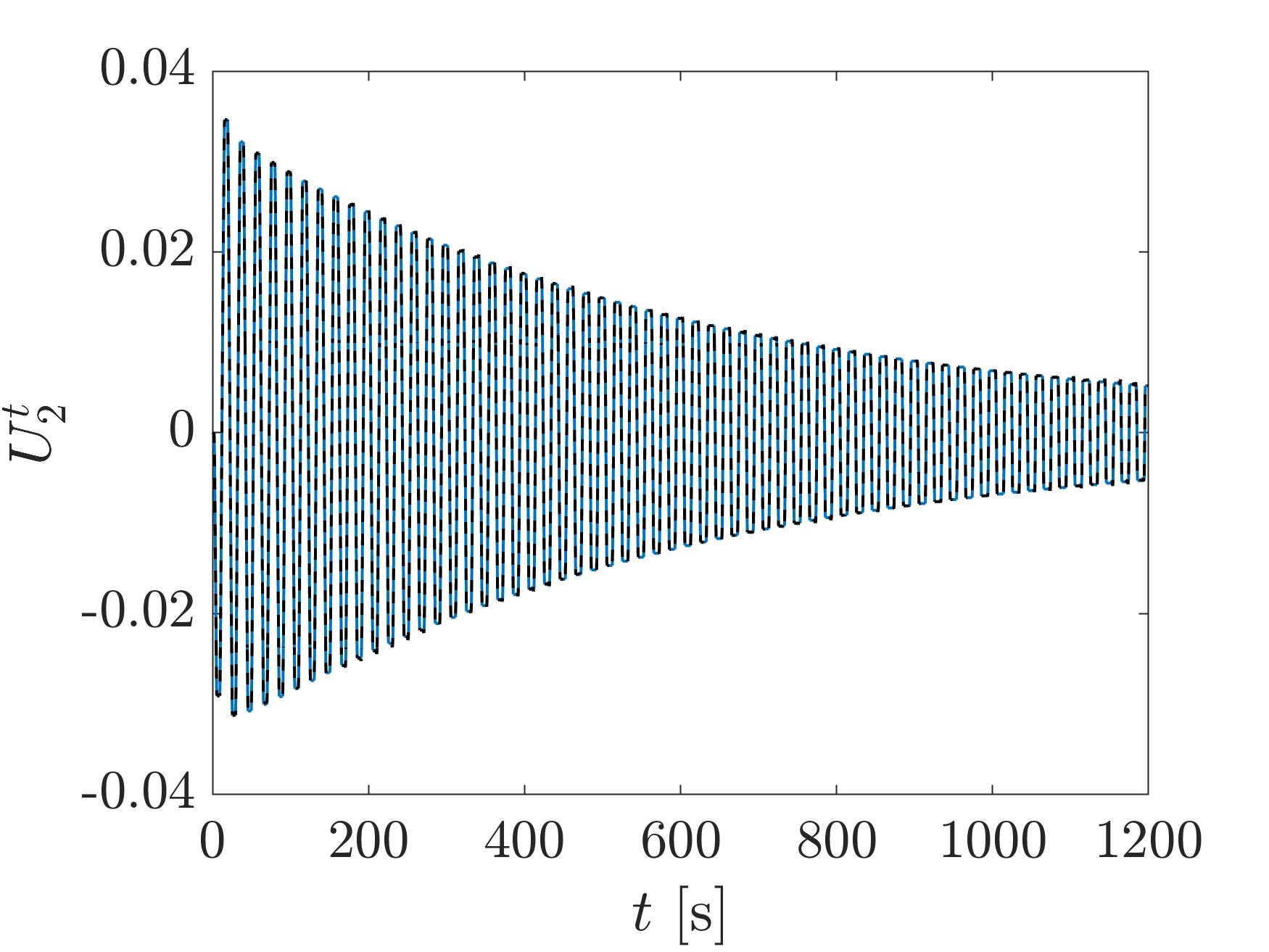}
		\end{subfigure}
		\begin{subfigure}{0.2\textwidth}
			\centering
			\includegraphics[trim={270 200 160 90},clip,width=1\textwidth]{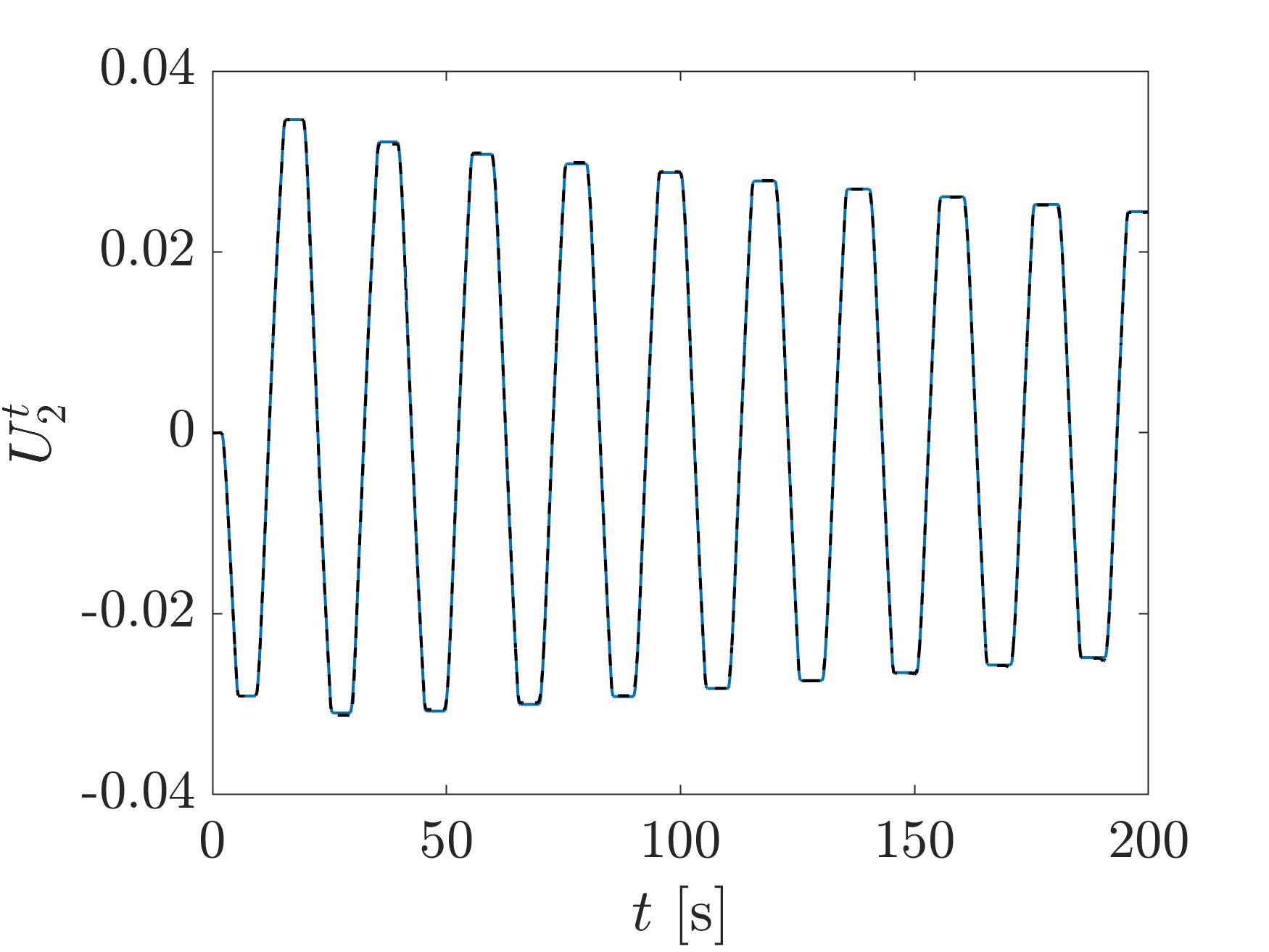}
		\end{subfigure}
		\begin{subfigure}{0.45\textwidth}
			\centering
			\includegraphics[width=1\textwidth]{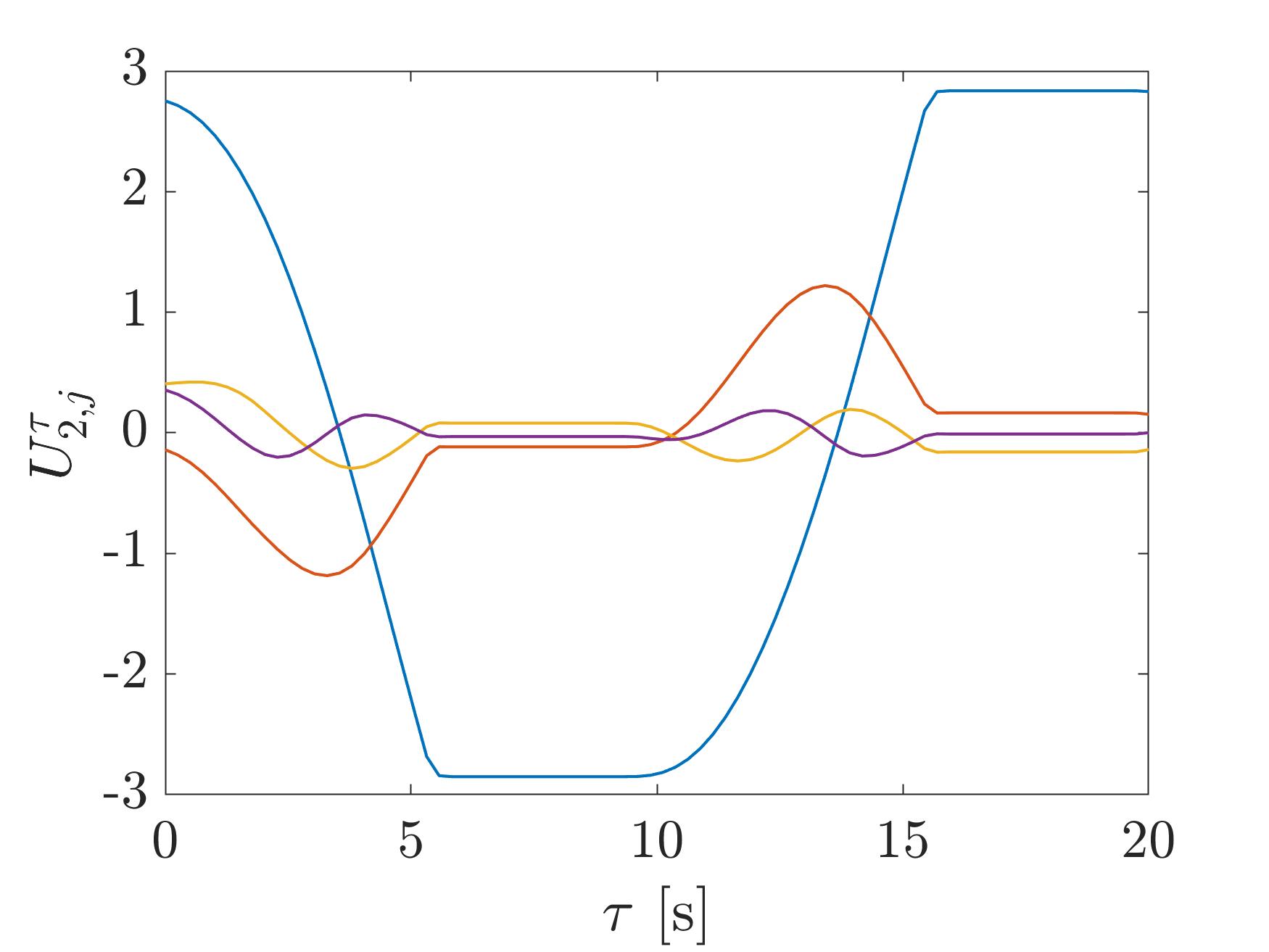}
		\end{subfigure}
		\begin{subfigure}{0.45\textwidth}
			\centering
			\includegraphics[width=1\textwidth]{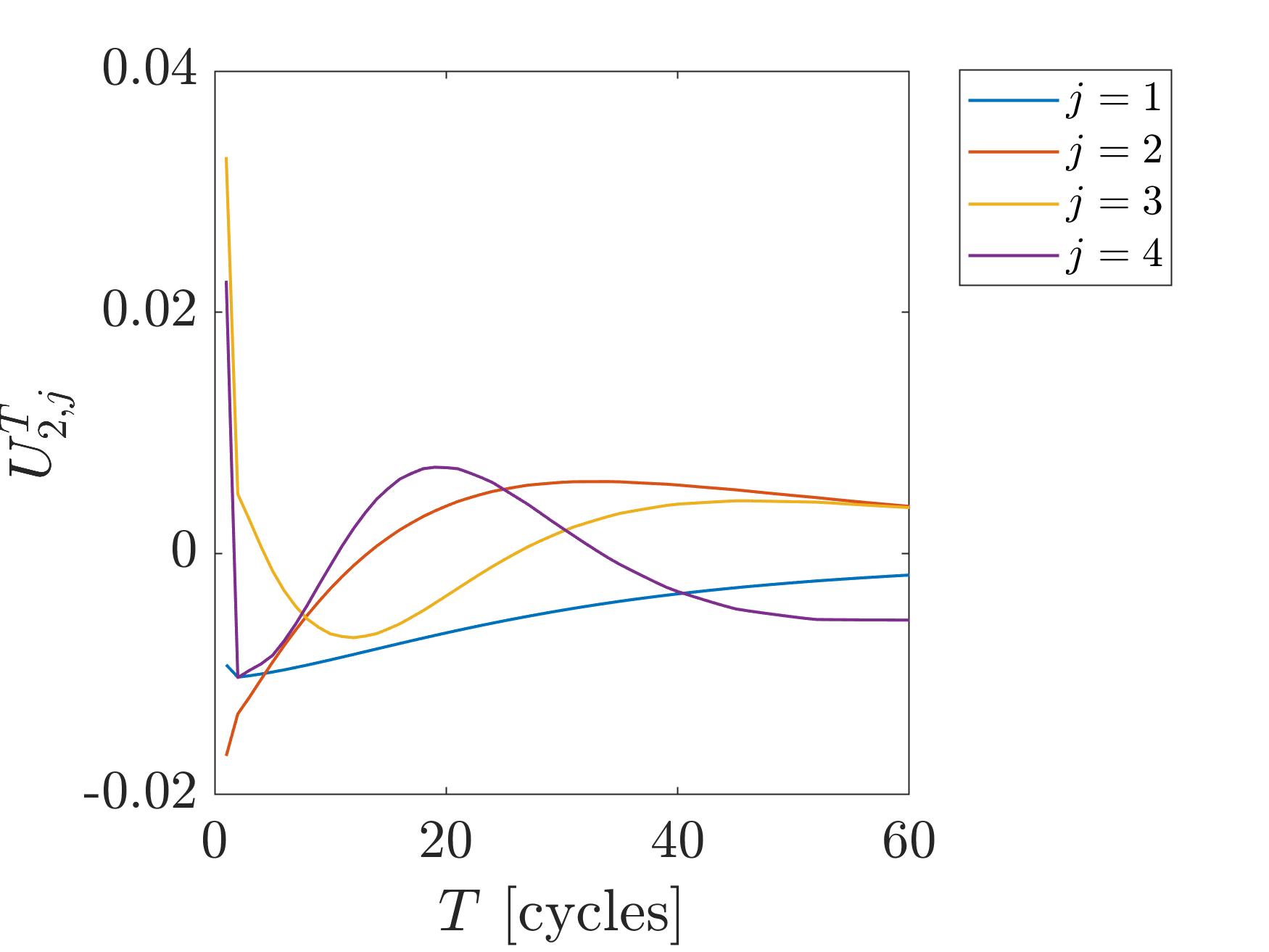}
		\end{subfigure}
		\caption{First four micro-macro modes of the multi-time decomposition of $U_2^t$.}
		\label{fig:mt-time-funcs}
	\end{figure}

	When employing the MT-PGD, each one of the signals in figure \ref{fig:time-modes-mtpgd-60} is approximated in terms of micro-macro submodes. For instance, the top of figure \ref{fig:mt-time-funcs} shows the second mode $U_2^t(t)$ computed via the PGD algorithm and by its multi-time counterpart MT-PGD (black dashed and blue lines superposed as shown in the zoomed figure). The images at bottom show the micro-time modes $\{U_{2,j}^{\tau}(\tau)\}_{j = 1}^{4}$ and macro-time ones $\{U_{2,j}^{T}(T)\}_{j = 1}^{4}$, respectively.
	
	As shown in figure \ref{fig:mt-time-funcs}, the micro-macro characterization is strongly physically consistent with the previously highlighted evolution of the time response. Indeed, the microscale functions capture the almost-periodicity of the function, through cyclic highly nonlinear patterns exhibiting fast dynamics. On the contrary, the dynamics is really slow along the macroscale, whose functions present an initial transitorial behavior, followed by a smooth evolution.
	
	Let us briefly comment about the choice of macropartitions (i.e., the choice of $\Delta T$) in the definition of the macroscale. As formerly discussed in \cite{pgd-multiscale}, the tensorial decomposition beyond the multi-time strategy makes this choice completely arbitrary, however results may be affected. As observed in \cite{pgd-multiscale}, if the choice is not physically meaningful, the convergence of the multiscale approximation may be exacerbated (more time submodes might be required to achieve convergence). 
		
	In order to achieve optimal convergence, a first concern is thus establishing a physically consistent decomposition of the time domain. For instance, referring to figure \ref{fig:micro-macro-mesh}, a wise solution could be choosing $\Delta T = k t_C$, with $k \in \mathbb{N}^+$, meaning that $\Delta T$ is a multiple of the external excitation period. Afterwards, the choice of $k$ may be driven by both physical and computational reasons. On the one hand, it may depend on what one aims at capturing along the fast scale (for instance, $k = 1$ corresponds to a full cycle response). On the other hand, the separated problems arising within the submodes computation are scaling with $N_\tau$ and $N_T$ for the micro and macro scales, respectively. Letting $N$ be the number of cycles, one gets $N_T = N/k$ and $N_\tau = N_t/N_T$, therefore the choice of $k$ affects their complexities.
	
	\subsection{Plate with edge crack}
	
\begin{figure}[H]
		\centering
		\begin{subfigure}{0.45\textwidth}
			\centering
			\includegraphics[width=1\textwidth]{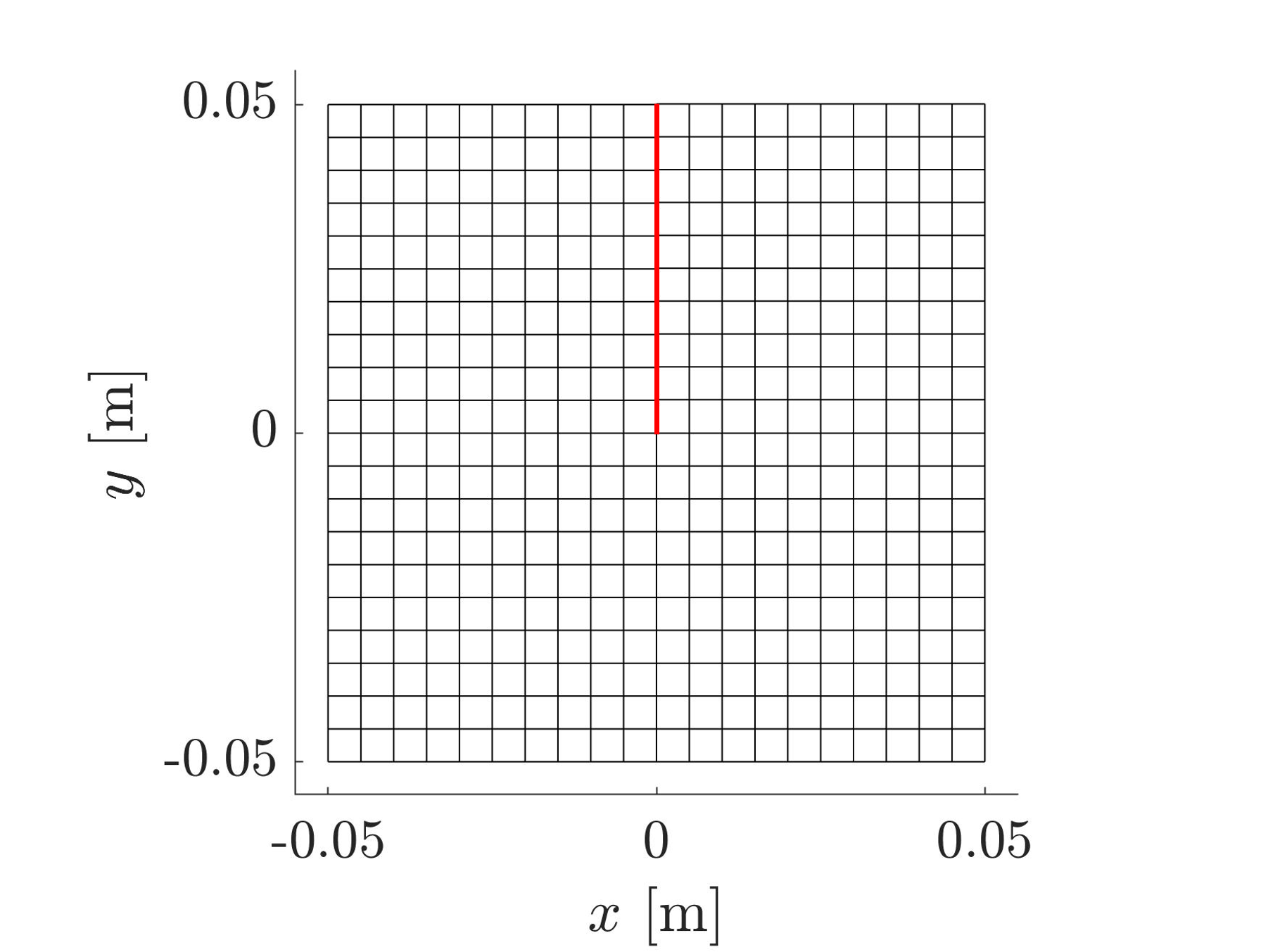}
		\end{subfigure}
		\begin{subfigure}{0.45\textwidth}
			\centering
			\includegraphics[width=1\textwidth]{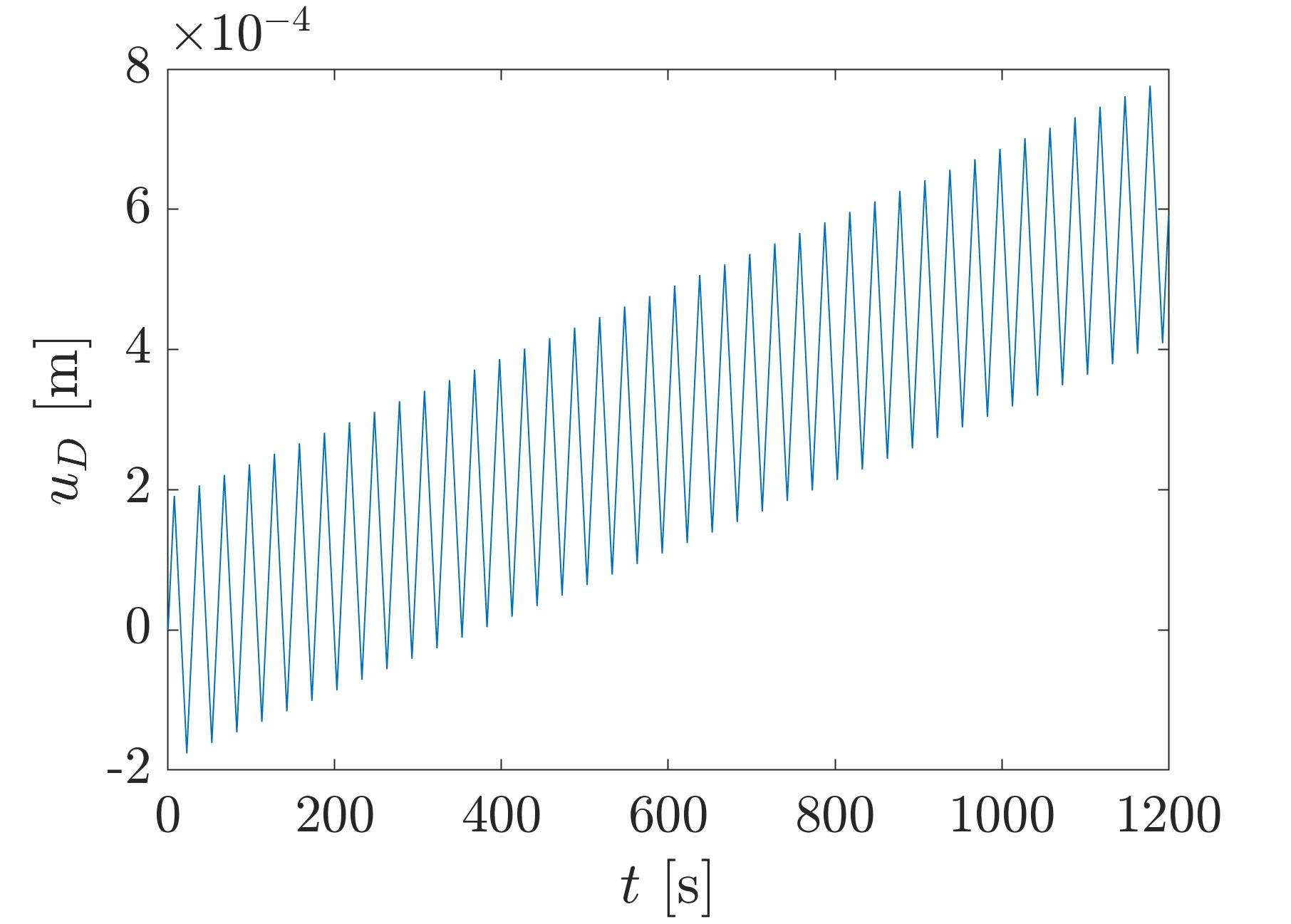}
		\end{subfigure}
		\caption{Discretized geometry (left) and imposed displacement (right).}
		\label{fig:test-case-plate}
	\end{figure}

	A second application case concerns a 2D plate of the same material, having an edge crack in the upper part, represented by the red segment in the left-side of figure \ref{fig:test-case-plate}. The simulation is, once again, controlled imposing a given displacement evolution to the right and left sides of the plate. Moreover, we consider a displacement imposed over 40 cycles with a given slope (no more centered in zero), as shows the right-side of figure \ref{fig:test-case-plate}. The spatial mesh consists of $N_e = 400$ quadrilateral elements and $N_x = 451$ nodes. The time interval is divided in $N_t = 3200$ times. A single cycle (load-unload-load) time has duration $T_1 = 30$ s, meaning that the final time is $T_f = 1200$ s.
	
\begin{figure}[H]
		\centering
		\begin{subfigure}{0.45\textwidth}
			\centering
			\includegraphics[width=1\textwidth]{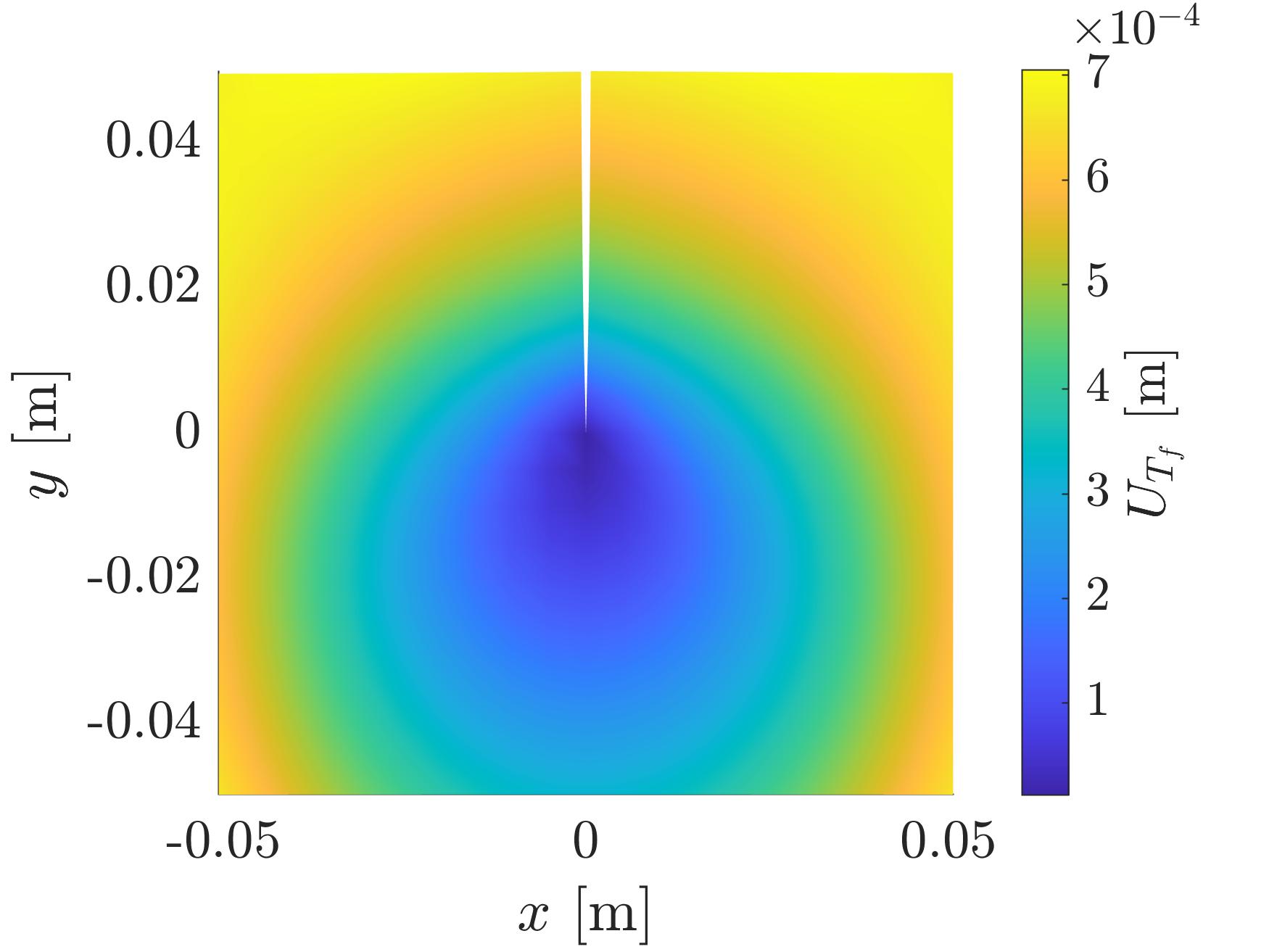}
		\end{subfigure}
		\begin{subfigure}{0.45\textwidth}
			\centering
			\includegraphics[width=1\textwidth]{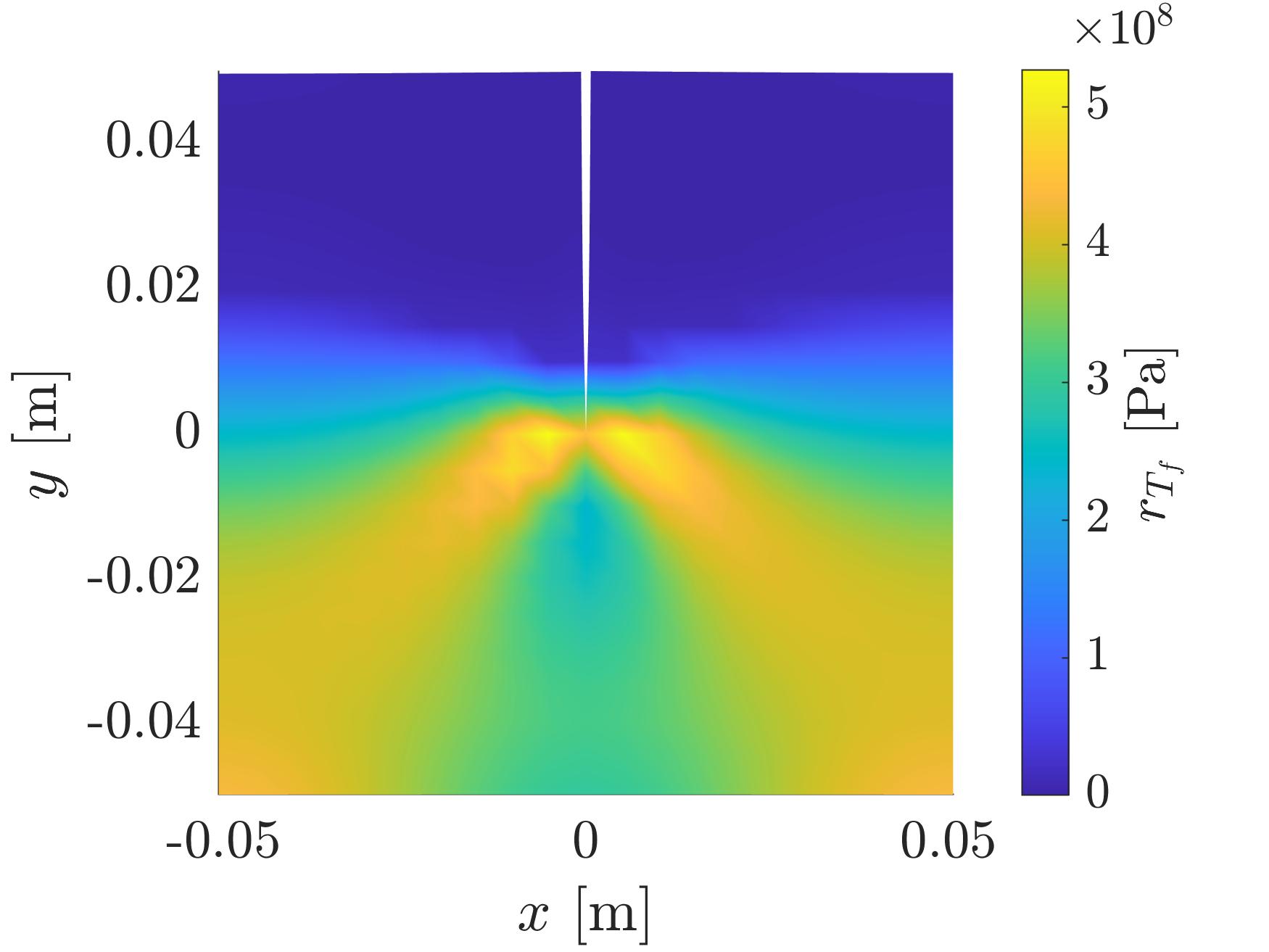}
		\end{subfigure}
		\caption{Displacement field (left) and isotropic hardening (right) at final time $T_f$.}
		\label{fig:test-case-results-plate}
	\end{figure}

	Figure \ref{fig:test-case-results-plate} gives the magnitude of the displacement field and the isotropic hardening function computed at the final time $T_f = 1200$ s.
	
\begin{figure}[H]
		\centering
		\begin{subfigure}{0.45\textwidth}
			\centering
			\includegraphics[width=1\textwidth]{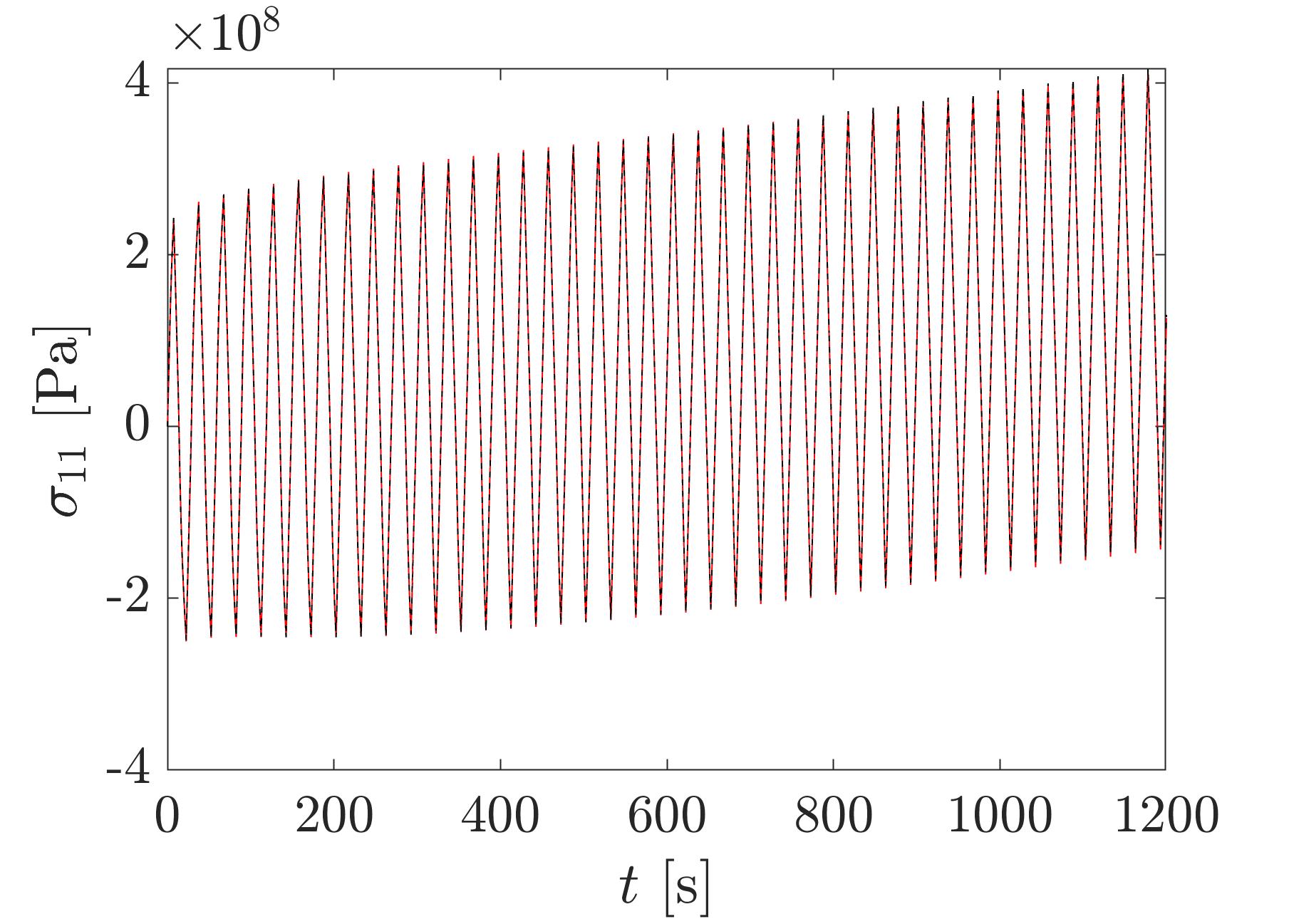}
		\end{subfigure}
		\begin{subfigure}{0.45\textwidth}
			\centering
			\includegraphics[width=1\textwidth]{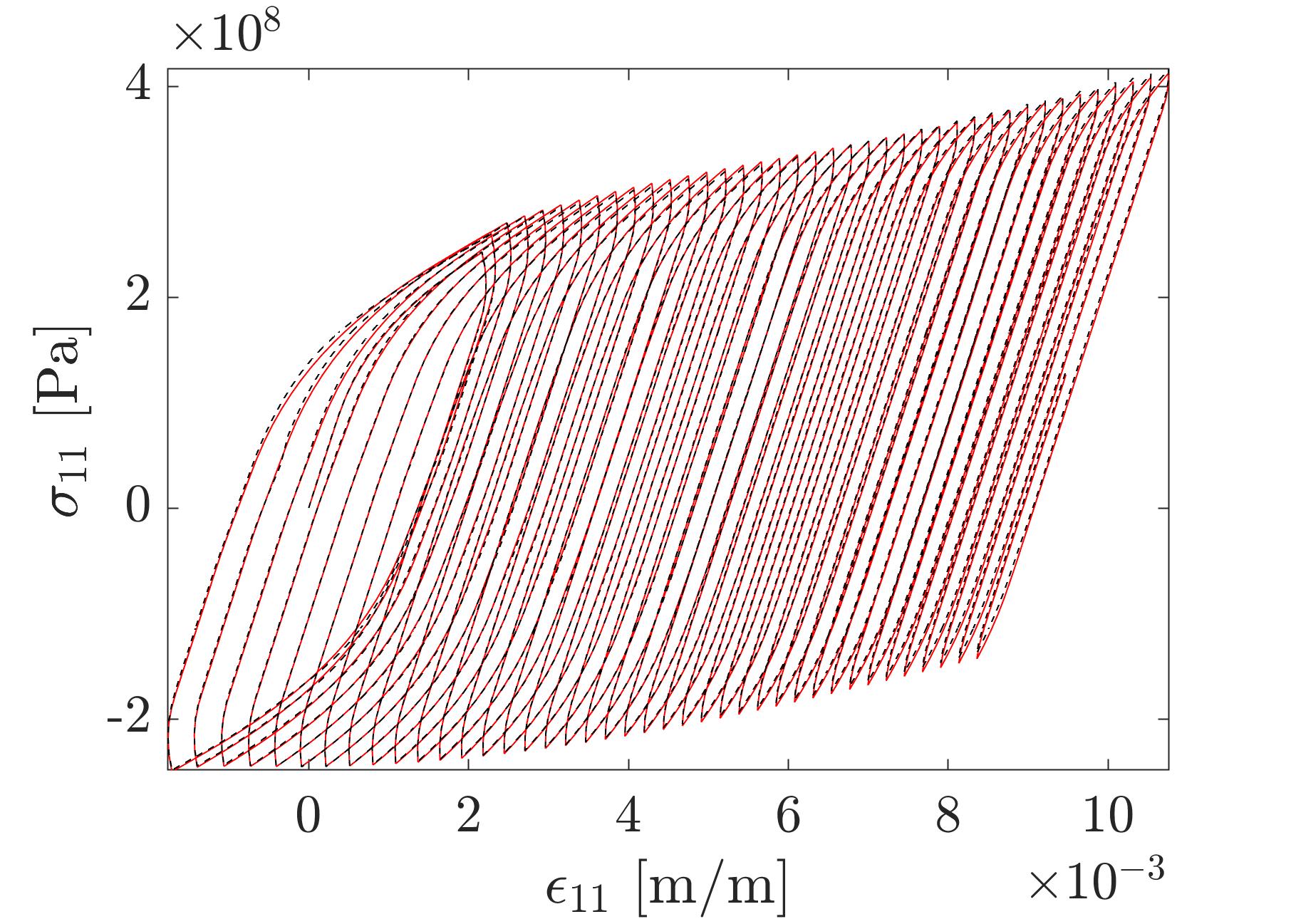}
		\end{subfigure}
		\caption{Stress-displacement curve (left) and hysteresis loop (right) in $(0,0)$. Red line: FE, black dashed line: PGD.}
		\label{fig:local-response-plate}
	\end{figure}

	Figure \ref{fig:local-response-plate} shows the comparison of the PGD results with a classical FE-based incremental algorithm (considering implicit integration of plasticity based on return-mapping) computed at the center of the specimen $(x_0, y_0) = (0,0)$.
	
\begin{figure}[H]
		\centering
		\includegraphics[width=1\textwidth]{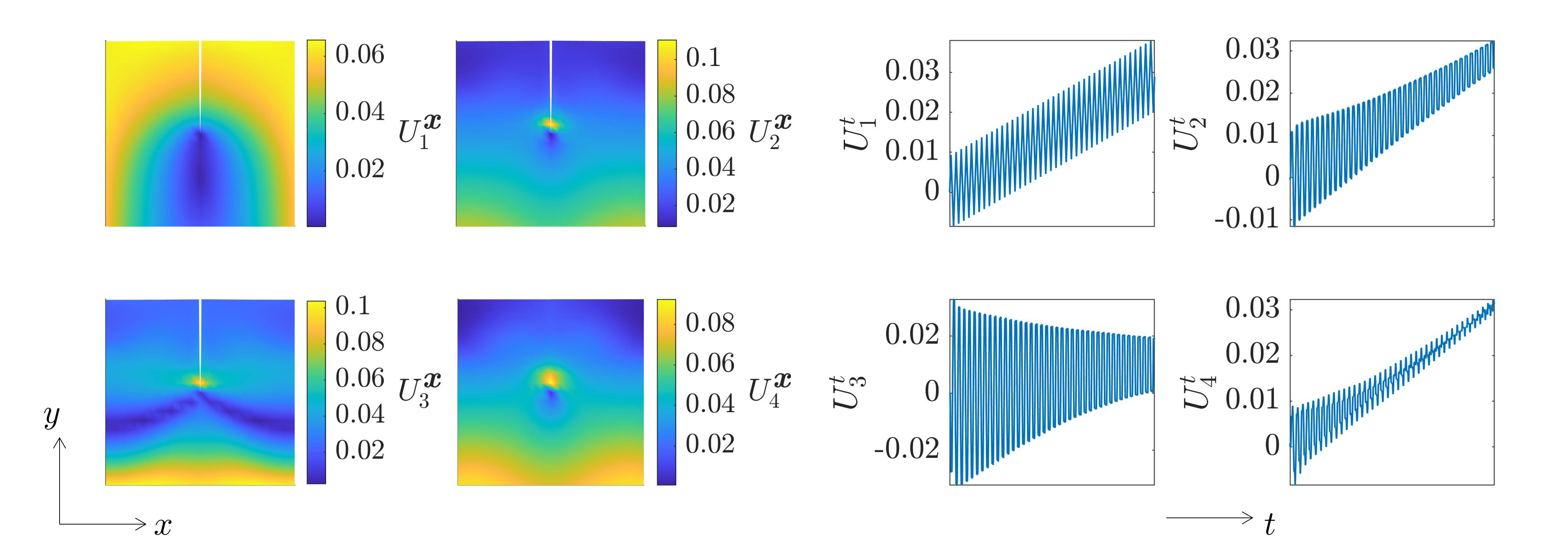}
		\caption{First four normalized PGD modes.}
		\label{fig:pgd-modes-plate}
	\end{figure}

	The first four normalized modes in space and time of the PGD approximation \eqref{eq:space-time-solution} are shown in figure \ref{fig:pgd-modes-plate}.
	
\begin{figure}[H]
		\centering
		\begin{subfigure}{0.45\textwidth}
			\centering
			\includegraphics[width=1\textwidth]{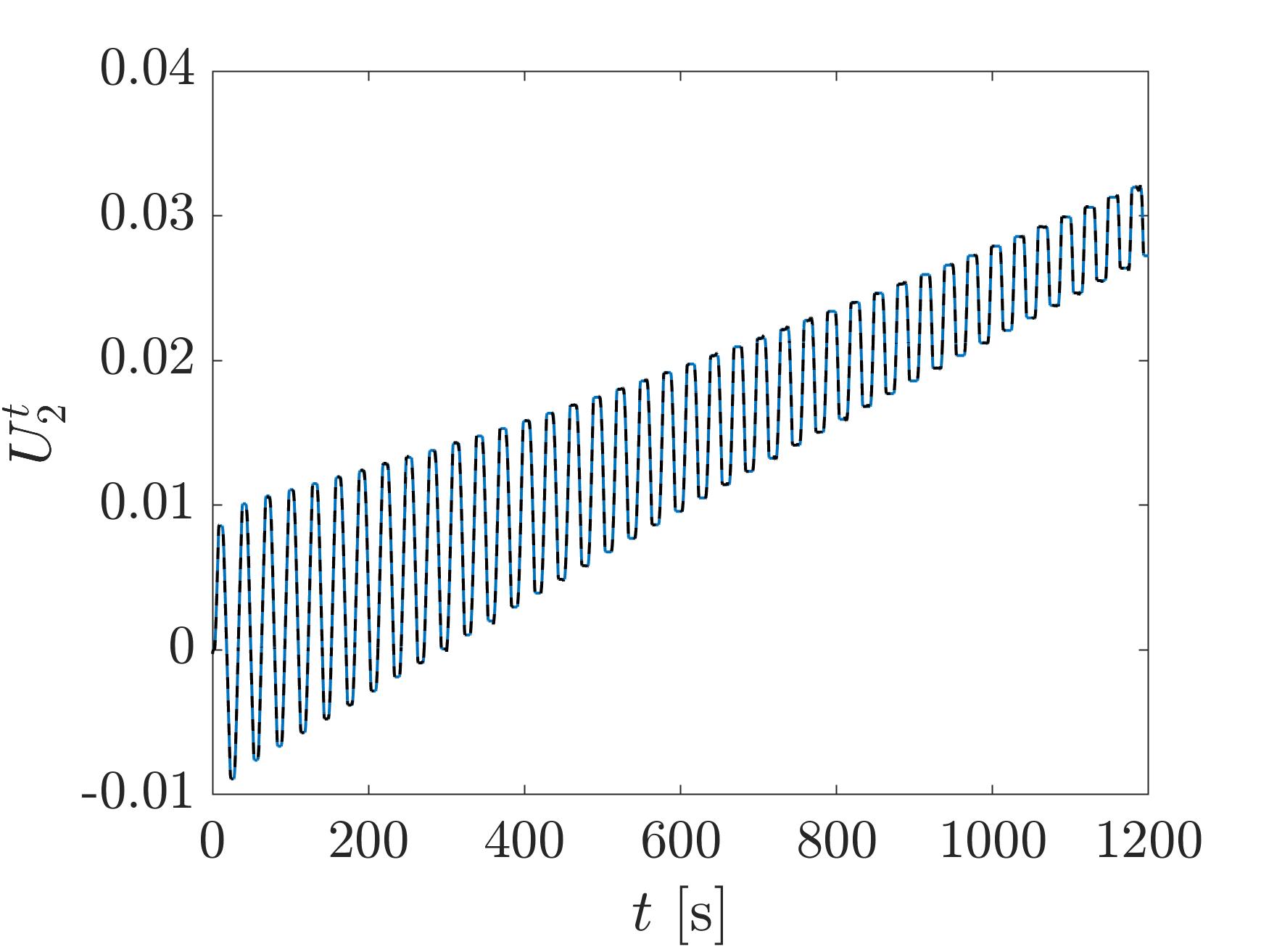}
		\end{subfigure}
		\begin{subfigure}{0.2\textwidth}
			\centering
			\includegraphics[trim={310 200 150 90},clip,width=1\textwidth]{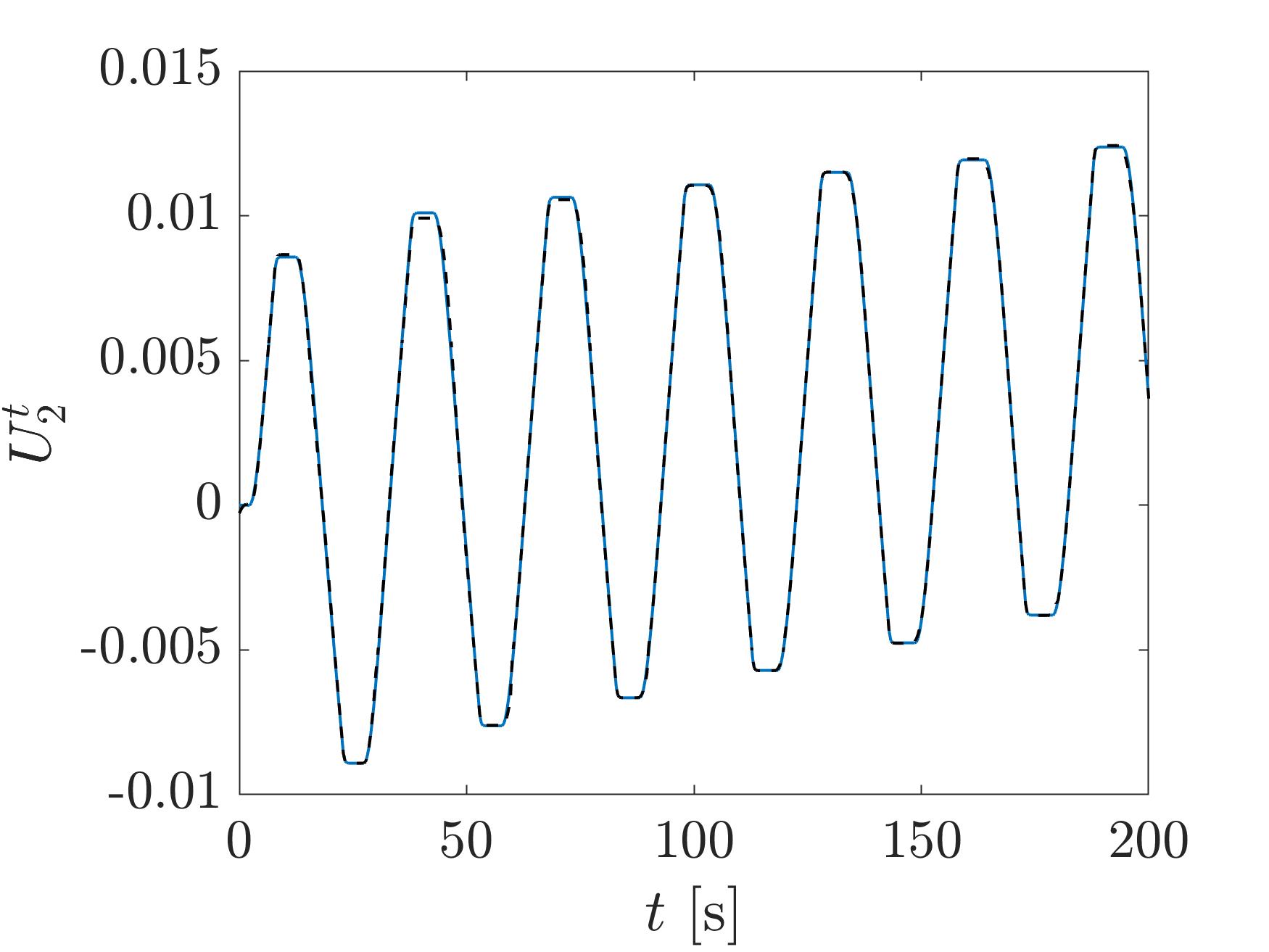}
		\end{subfigure}
		\begin{subfigure}{0.45\textwidth}
			\centering
			\includegraphics[width=1\textwidth]{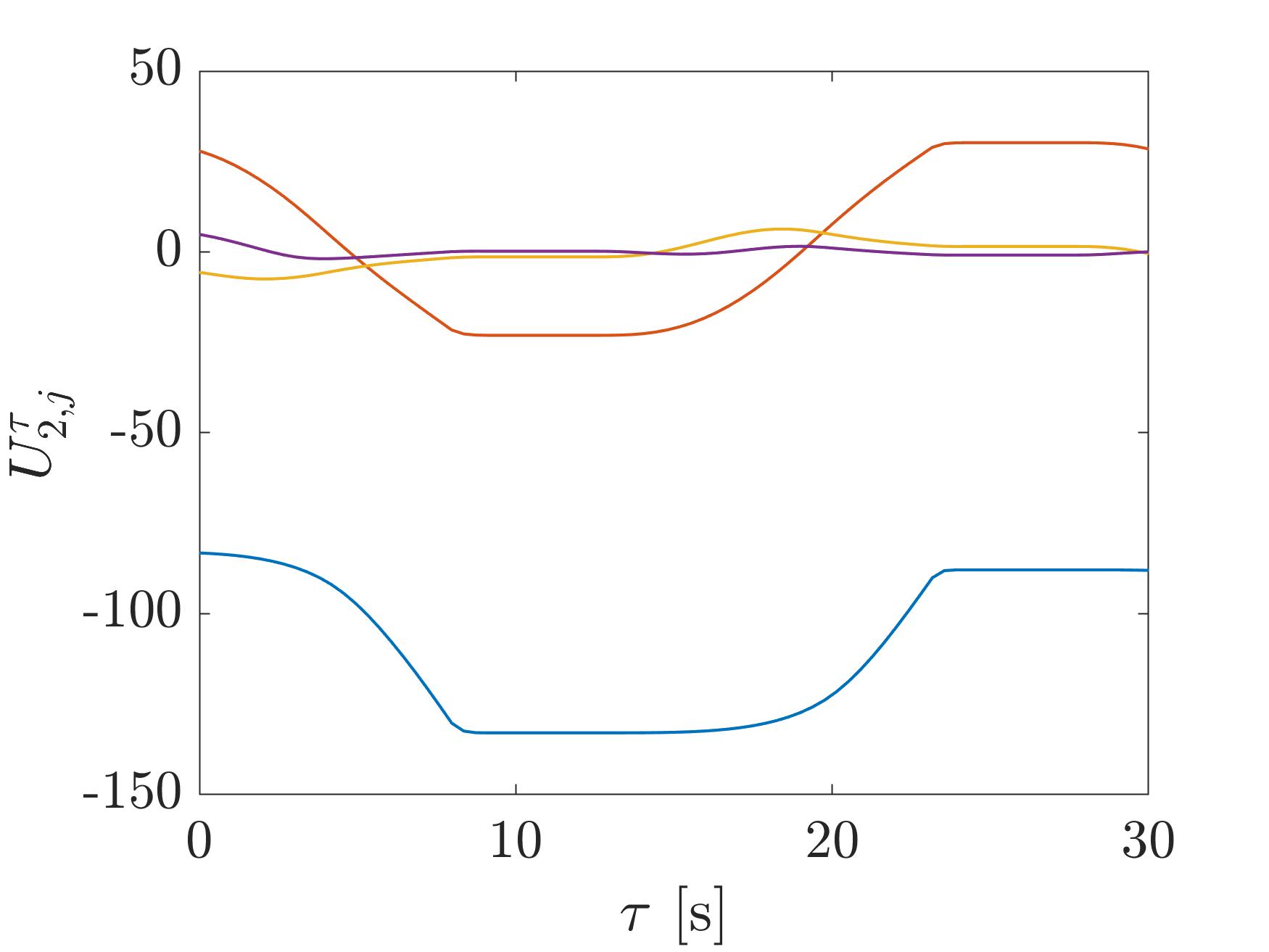}
		\end{subfigure}
		\begin{subfigure}{0.45\textwidth}
			\centering
			\includegraphics[width=1\textwidth]{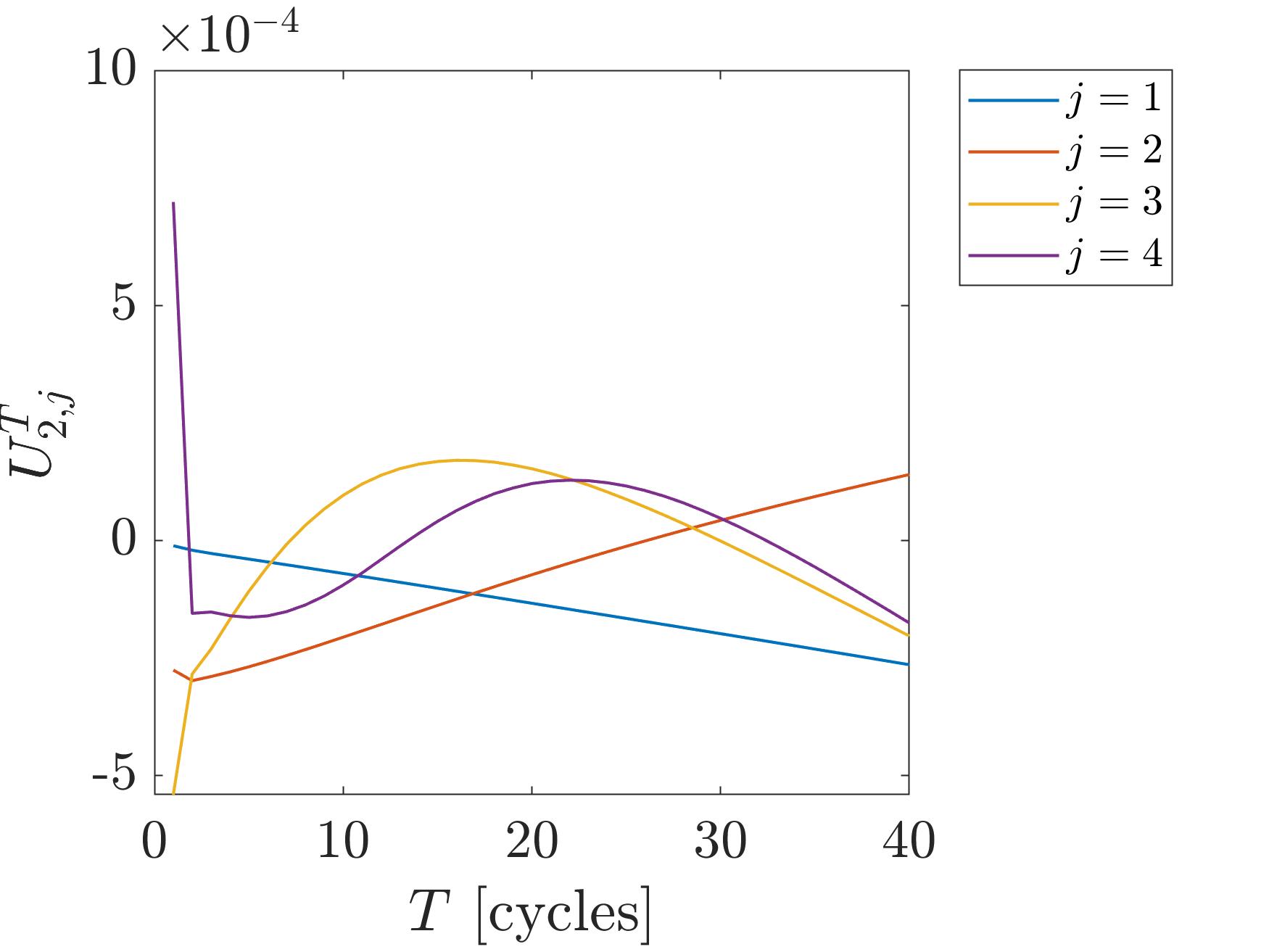}
		\end{subfigure}
		\caption{First four micro-macro modes of the multi-time decomposition of $U_2^t$.}
		\label{fig:mt-time-funcsplate}
	\end{figure}

	The time modes in figure \ref{fig:pgd-modes-plate} clearly exhibit a multiscale behavior, easily recognized when employing the MT-PGD. As for the previous example, the number of macrodofs corresponds to the number of cycles $N_T = 40$, while the microscale has $N_\tau = 80$ times spanning the first macro-interval (in high-cycle analyses, a larger $\Delta T$ could be preferred to reduce the complexity of the macroscale problem). The multi-time decomposition of the second PGD time mode $U^t_2$ is illustrated in figure \ref{fig:mt-time-funcsplate}. We can observe that the repeating patterns are identified by the microscale modes, while their stabilization is slower and retrieved through the macroscale modes, which track the almost linear trend at large scale.

    \section{Conclusions}
    \label{sec:conclusion}
    As an extension of \cite{pgd-multiscale}, this study successfully accomplished the computation of time-separated solutions in the nonlinear setting of cyclic elasto-plasticity. 
    
    Within the PGD-based time multiscale procedure, the time response is computed along two separated time scales, the micro and the macro one. Such scales are defined as newly independent coordinates, while the full scale is recovered via their tensor product. The study shows physically consistent results for the elasto-plastic response under cyclic loading. Indeed, delegating a whole cycle to the microscale, highly nonlinear patterns are observed over the fast scale, while a smooth and slow evolution is captured by the macroscale. This makes the macroscale characterization particularly attractive for long-time horizon analyses, such as aging and fatigue \cite{cyclic-plasticity-2}.
  	
  	In \cite{pgd-multiscale-2}, authors propose an efficient algorithm to optimally decompose complex signals (involving many frequencies) in a fast and slow scale. Prior to the multi-time approximation, the algorithm from \cite{pgd-multiscale-2} may be applied to the external excitation of the problem at hand. In such a way, an optimal decomposition of the time axis may be established apriori (i.e., the best value of the macroscale step size $\Delta T$ is determined), guaranteeing the optimal convergence of the multi-time PGD procedure. Works in progress are dealing with this topic. 
	
	For what concerns the nonlinear character of the problem, the linearization procedure is based on solving over the full space/time domain the elastic problem and enforcing the plastic contribution to the right-hand-side. The integration of plasticity is performed through the elastic predictor/return-mapping algorithm \cite{plasticity-1, plasticity-2}. This can be viewed as a nonlinear operator $\mathcal{N}$ acting on the total strain tensor and on the effective plastic strain up to the final time $T_f$, that is
	\begin{equation}
		\label{eq:non-linear-term-hist}
		\ctens{\varepsilon}^{p, (l - 1)} =  \mathcal{N}(\ctens{\varepsilon}^{(l - 1)}, \bar{\varepsilon}^p_{T_f}).
	\end{equation} 
	Specifically, the evaluation of \eqref{eq:non-linear-term-hist} requires the reconstruction over the whole past history since
	\begin{equation}
		\bar{\varepsilon}^p_{t} = \int_0^{t} \sqrt{\frac{2}{3} \dot{\ctens{\varepsilon}}^p : \dot{\ctens{\varepsilon}}^p } \dd{s}.
	\end{equation}
	In this sense, so far, the proposed methodology shares the usual drawbacks of standard techniques in computational plasticity since the integration of plasticity does not benefit from the multi-time format of the time scheme. 
	
	Current research is also focusing on this matter. In particular, authors are exploiting the multiscale representation to build an efficient data-driven model of the elasto-plastic constitutive relation. This would allow real-time evaluations of \eqref{eq:non-linear-term-hist}, enabling the direct simulation of inelasticity also in long-term scenarios, such as high-cycle fatigue.
	
	As a final comment, the method benefits of the usual advantages entailed by PGD-based procedures. For instance, model parameters and loading conditions can be treated as problem extra-coordinates, enabling the fast computation of multi-parametric solutions \cite{PGD-param-0, PGD-param-1, PGD-param-2}. Moreover, the further time separation guarantees, when solving the linearized problem, the same operational and memory savings discussed in \cite{pgd-multiscale}. 

    \phantomsection
    \section*{References}
    \addcontentsline{toc}{section}{References}
    \printbibliography[heading=none]

\end{document}